\documentclass[conference]{IEEEtran}
\let\labelindent\relax
\usepackage{amsmath,amsthm,amsfonts} 
\usepackage{algorithmic}
\usepackage{graphicx}
\usepackage{textcomp}
\usepackage{xcolor}
\usepackage{multirow}
\usepackage{caption}
\usepackage{xspace}
\usepackage{enumitem}
\usepackage{booktabs}
\usepackage{makecell}
\usepackage{bm}
\usepackage{balance}
\usepackage{bbding}
\begin{document}

\newcommand{\etal}{\textit{et al}}
\newcommand{\eg}{\emph{e.g.,}\xspace}
\newcommand{\ie}{\emph{i.e.,}\xspace}
\newcommand{\resp}{\textit{resp}.\xspace}
\newcommand{\ours}{RAPIER\xspace}
\newcommand{\Cor}{\textsf{Corrector}\xspace}
\newcommand{\Gen}{\textsf{Generator}\xspace}
\newcommand{\todo}[1]{{\textcolor[RGB]{174, 76, 207}{(Todo: #1)}}}
\newcommand{\first}{\textsf{(i)}\xspace}
\newcommand{\second}{\textsf{(ii)}\xspace}
\newcommand{\third}{\textsf{(iii)}\xspace}
\newcommand{\xinhao}[1]{{\textcolor[RGB]{44, 115, 210}{(Xinhao: #1)}}}
\newcommand{\qing}[1]{#1}
\newcommand{\new}[1]{{\textcolor[RGB]{174, 76, 207}{#1}}}

\title{\fontsize{23pt}{22pt}\selectfont Low-Quality Training Data Only? A Robust Framework for Detecting Encrypted Malicious Network Traffic}
\author{
    \IEEEauthorblockN{
        Yuqi Qing\IEEEauthorrefmark{4}\IEEEauthorrefmark{1}\thanks{\IEEEauthorrefmark{1}The first two authors contributed equally to this paper.},
        Qilei Yin\IEEEauthorrefmark{2}\IEEEauthorrefmark{1},
        Xinhao Deng\IEEEauthorrefmark{4},
        Yihao Chen\IEEEauthorrefmark{5},
        Zhuotao Liu\IEEEauthorrefmark{4},
        Kun Sun\IEEEauthorrefmark{3},
        Ke Xu\IEEEauthorrefmark{5},
        Jia Zhang\IEEEauthorrefmark{4},
        Qi Li\IEEEauthorrefmark{4},
    }
    \IEEEauthorblockA{
        \IEEEauthorrefmark{4}Institute for Network Sciences and Cyberspace \& BNRist, Tsinghua University, \IEEEauthorrefmark{2}Zhongguancun Laboratory
    }
        
    \IEEEauthorblockA{
        \IEEEauthorrefmark{5}Department of Computer Science, Tsinghua University, \IEEEauthorrefmark{3}George Mason University
    }
    \IEEEauthorblockA{
        \{qyq21, dxh20, yh-chen21\}@mails.tsinghua.edu.cn,
        yinql@zgclab.edu.cn,
    }
    \IEEEauthorblockA{
        \{zhuotaoliu, xuke, zhangjia2017, qli01\}@tsinghua.edu.cn,
        ksun3@gmu.edu
    }
}

\IEEEoverridecommandlockouts
\makeatletter\def\@IEEEpubidpullup{6.5\baselineskip}\makeatother

\IEEEpubid{
    \parbox{\columnwidth}{
    Network and Distributed System Security (NDSS) Symposium 2024\\
    26 February - 1 March 2024, San Diego, CA, USA\\
    ISBN 1-891562-93-2\\
    https://dx.doi.org/10.14722/ndss.2024.23081\\
    www.ndss-symposium.org
}
\hspace{\columnsep}\makebox[\columnwidth]{}}

\maketitle

\begin{abstract}
Machine learning (ML) is promising in accurately detecting malicious flows in encrypted network traffic; however, it is challenging to collect a training dataset that contains a sufficient amount of encrypted malicious data with correct labels. When ML models are trained with low-quality training data, they suffer degraded performance. In this paper, we aim at addressing a real-world low-quality training dataset problem, namely, detecting encrypted malicious traffic generated by continuously evolving malware. We develop \ours that fully utilizes different distributions of normal and malicious traffic data in the feature space, where normal data is tightly distributed in a certain area and the malicious data is scattered over the entire feature space to augment training data for model training. \ours includes two pre-processing modules to convert traffic into feature vectors and correct label noises. We evaluate our system on two public datasets and one combined dataset. With 1000 samples and 45\% noises from each dataset, our system achieves the F1 scores of 0.770, 0.776, and 0.855, respectively, achieving average improvements of 352.6\%, 284.3\%, and 214.9\% over the existing methods, respectively. Furthermore, We evaluate \ours with a real-world dataset obtained from a security enterprise. \ours effectively achieves encrypted malicious traffic detection with the best F1 score of 0.773 and improves the F1 score of existing methods by an average of 272.5\%. 

\end{abstract}

\section{Introduction}
\label{sec:introduction}

\qing{Network-based intrusion detection approaches~\cite{bartos2016optimized, fu2021, fu2022ton, fu2023, jan2020odds, nelms2013execscent, paxson1999bro, roesch1999snort, wang2017detecting, zhou2023netbeacon} have been extensively developed to detect malicious traffic in different networks~\cite{site-network-cisco, site-network-cloudflare,site-network-azure}.} 
As more malware samples start to use encryption protocols to hide traffic content, the traditional detection methods focusing on analyzing plaintext payloads are obsoleted. 
\qing{Learning-based methods~\cite{ anderson2016cisco, anderson2017cisco, fu2022encrypted, liu2018mampf, liu2019fsnet, tegeler2012botfinder, wang2016trafficav, xie2023rosetta, zhang2018HoMonit, zheng2020learning} advanced malicious behavior detection by analyzing encrypted network flows.}
These designs are typically supervised, heavily relying on a training dataset that contains high-quality data samples to build accurate and robust detection models.

It is non-trivial to collect high-quality training data.  
First, collecting time-sensitive malware data is difficult. The typical approach is to execute malware samples captured in real-world cyberspace in controlled sandboxes and collect the generated traffic~\cite{miramirkhani2017spotless,willems2007toward}. However, since the malware evolves consistently, the captured malware samples have insufficient time sensitivity.
Second, it is difficult to label the collected data in practice. 
\qing{Data labels predicted by malware detection services are not always reliable~\cite{xu2021differential}. For instance, labels assigned to malware by Virustotal may vary in the reports published in different years. Moreover, the cost of labeling data manually is non-negligible~\cite{guerra2022datasets,xu2021differential}.}
The current labeling approach results in potential label noises in collected datasets. Unfortunately, the encryption protocols (\eg SSL/TLS)  prevent us from manually correcting these noises.
Therefore, the training dataset collected in reality is typically limited in both quality and size. 

\qing{Prior art~\cite{frenay2013classification, patrini2017making,shorten2019survey,song2022learning,wang2006suppressing,wang2020generalizing} shows that a limited number of training samples or label noises of training samples can lead to the degradation of model generalizability on new data.} \qing{The issue becomes worse when the limited training samples have label noises, \ie low-quality training data.}
\qing{However, existing techniques like data augmentation~\cite{ferdowsi2019GANIoT, jan2020odds, lim2018doping, zenati2018efficientGAN} and robust ML models against label noises~\cite{chen2015webly, goldberger2016training, han2018coteaching, hendrycks2018using, liang2021FARE, malach2017decoupling, patrini2017making, song2019selfie, sukhbaatar2014training, xia2020part, xu2021differential, yao2020dual, yu2019does} cannot address the issue.}
Specifically, the data augmentation methods synthesize new data based on the distributions of existing training data. However, label noises will confuse the distributions of different categories, resulting in new data being synthesized from an incorrect distribution, which could create more label noises. 
Further, to correct the impacts of label noise on model training, existing art relies on strong assumptions and prior knowledge~\cite{chen2015webly, goldberger2016training, hendrycks2018using, patrini2017making, sukhbaatar2014training, xia2020part, yao2020dual} (\eg the probability of a sample being mislabeled) or a large-sized training set that can reveal each sample's intrinsic characteristics~\cite{han2018coteaching, malach2017decoupling, song2019selfie, xu2021differential, yu2019does}. None of these prerequisites are satisfied in our problem. 
\qing{Recent encrypted traffic classification methods~\cite{lin2022bert, zhao2022mt} improved the performance under limited labeled training data by transferring knowledge from additional large-scale unlabeled training data.} However, collecting and pre-processing such a large-scale dataset is expensive. It may also increase the risk of privacy leakage, \ie an attacker can infer attributes of training data from a trained ML model~\cite{jia2018attriguard}.

In this paper, we propose a novel encrypted malicious traffic detection system \ours that is robust to low-quality training data. The high-level idea is to leverage the difference in distribution between benign and malicious traffic data to estimate the possible location of each type of data. Since our system does not depend on the correctness of the sample labels or the amount of training data, it can be trained on a low-quality dataset that contains label noises.
Since normal behaviors are typically more representative and consistent than malicious ones, benign data tends to exhibit a denser distribution than malicious data. 
Thus, we can infer the true labels for training samples based on their data distributions. Armored attackers may imitate normal behaviors by generating malicious data with a distribution similar to that of normal data over time. 
In this case, we synthesize new training data located in the possible distribution regions of new malicious data to improve the generalizability of our models. 

\ours is powered by three tightly coupled components.
First, we propose a novel feature extraction module to convert raw encrypted network traffic into feature vectors representing fine-grained behaviors. Based on an improved auto-encoder architecture, our feature extraction module works in an unsupervised manner to prevent incorrect labels from contaminating the feature vectors. 
Second, we design a distribution-aware label noise correction module to infer the true labels of the original training samples. It estimates the distribution of training data via an auto-regressive generative model, re-labels the training data exhibiting the most obvious distribution characteristics, and infers the labels of remaining data through ensemble learning. 
Third, we develop a new data augmentation module to synthesize new training data. With the label-corrected training data, this module selects the possible distribution regions of new malicious data and applies an improved Generative Adversarial Network (GAN) model to generate new training data that is located in these target regions. The synthetic data is combined with the label-corrected data for training. In summary, our contributions are as follows:

\begin{itemize}[leftmargin=*]
    \item We develop a system called \ours to detect encrypted malicious traffic in the case that the amount of training data is limited and non-negligible label noises exist. 
    It is the first malware traffic detection system that simultaneously overcomes the challenges of both training data insufficiency and label noises.
    \item We implement \ours and perform extensive evaluations based on two public datasets and one combined dataset. With only 1000 training samples (\ie 500 malicious and 500 normal ones) and 45\% noise ratio, \ours effectively achieves the F1 scores of 0.770, 0.776, and 0.855 on the three datasets, respectively, achieving average improvements of 352.6\%, 284.3\%, and 214.9\% over the existing methods.
    \item We evaluate \ours with a real-world dataset collected by a security enterprise. It achieves the best label correction performance when the noise ratios vary from 20\% to 45\%, consistently reducing the noise ratios to less than 4.3\%. Furthermore, it can effectively achieve encrypted malicious traffic detection with the best F1 score of 0.773 and improve the F1 score of existing methods by an average of 272.5\%.
    
\end{itemize}

\section{Background}
\label{sec:background}

\subsection{Data Augmentation}
Data augmentation techniques can effectively increase the size of the training data set without explicitly collecting new data. 
The most common data augmentation strategy is to oversampling, \ie replicate samples from the minority class. In particular, the new training samples can be randomly sampled from minority class examples~\cite{japkowicz2000class} or be synthesized based on random combinations of existing training samples and their nearest neighbors~\cite{chawla2002smote, han2005BorderlineSMOTE,he2008adasyn}. 
As a result, the ML model may overfit the training set due to the limited diversity of the data. 
\qing{Recently, the Generative Adversarial Network (GAN)~\cite{goodfellow2014GAN} has been developed and widely applied for data augmentation in various areas like image classification~\cite{frid2018synthetic}.} A typical GAN consists of two major components: the \textit{generator} learns a specific data distribution to generate new data and the \textit{discriminator} tries to distinguish the newly generated data from the original training data. Powered by an adversarial training framework, the generator may learn the original training data distribution and then create a variety of new training data that conform to this distribution for the purpose of augmenting the limited training data. However, all those methods are not resilient to label noises. The falsely labeled training samples will confuse the data distribution of different categories, generating more label noises. Thus, the existing data augmentation methods cannot handle the low-quality issue of encrypted training data.

\subsection{Robust Machine Learning Models}

Robust machine learning models aim to mitigate the impacts of incorrectly labeled training data on the generalizability of the models. In particular, several models~\cite{ghosh2017robust,lyu2019curriculum,wang2019symmetric,zhang2018generalized} utilize robust loss functions to achieve the same misclassification probability for the truly labeled and incorrectly labeled training data. They typically rely on one assumption that a sample in one category will be mislabeled into other categories with equal probability; however, this assumption does not always hold in practice. 
Other models~\cite{hendrycks2018using,patrini2017making} apply the label transition matrix, which records the probability of a category being mislabeled into another category, to correct loss values impacted by noise labels. However, the label transition matrix is usually unknown prior. Recent studies~\cite{arazo2019unsupervised, han2018coteaching,malach2017decoupling,song2019selfie} address the issue by automatically selecting incorrectly labeled samples from the noisy training set based on their properties, \eg, the loss value of a mislabeled sample is higher than that of truly labeled ones in the training stage. However, these properties may not hold when the number of training samples decreases and significantly constrain these methods.  

\section{Problem Statement}
\label{sec:problem}

This paper aims to develop a system that can detect malware infections within an internal network, \eg a campus or an enterprise network, by identifying the encrypted malicious traffic generated by malware. 
The detection system is deployed at the gateway of the intranet to monitor the outgoing traffic of all internal hosts simultaneously. In order to deploy the system, a network administrator needs to collect and label encrypted network traffic in the internal network to create a training set. However, due to the scarcity of malware samples and encrypted traffic payloads, the training set is usually of low quality, \ie the amount of training data is limited and the data is with non-negligible label noises. Since training sample collection requires manual inspection, the number of training samples is much smaller than the testing data and the samples may be wrongly labeled.

Formally let $\left ( x_i,y_i \right )$ be the pair of an encrypted sample $x_i$ (\eg a flow or a session between two hosts) and its true label $y_i \in \left \{ 0,1 \right \}$, where $0$ or $1$ represents a normal or a malicious one, respectively. The inputs of a detection system are a low-quality training set $\widetilde{D}_{train}= \left \{ \left ( x_i,\widetilde{y}_i \right )  \right \}_{i=1}^{N} $ and a testing set $D_{test}=\left \{ x^{test} \right \} $, where $N \ll \left | D_{test} \right | $ and $\widetilde{y}_i$ is a \emph{noisy} label that may be inconsistent with $y_i$. Our goal is to accurately infer the label of $x^{test}$ by using $\widetilde{D}_{train}$.

\qing{Our goal is to address the limitations of existing encrypted malicious traffic detection methods~\cite{anderson2016cisco, anderson2017cisco, fu2022encrypted, lin2022bert, liu2018mampf, liu2019fsnet, tegeler2012botfinder, wang2016trafficav, xie2023rosetta,zhang2018HoMonit,zhao2022mt,zheng2020learning}, and develop an encrypted malicious traffic detection system that is applicable in more realistic scenarios.}
First,  we consider two problems in the training data, namely, insufficient data and non-negligible label noises, whereas prior works only try to solve one problem. Second, our system does not require additional data to improve detection performance, except for a low-quality training set. 
\qing{Recent works~\cite{lin2022bert, zhao2022mt} utilize extra large-scale unlabeled training sets to transfer the knowledge for detection.}
However, collecting such large-scale traffic for training is time-consuming and may incur privacy leakage (\eg an attacker can infer attributes of training data from a trained ML model~\cite{jia2018attriguard}).
We do not consider the scenario in which malicious traffic exhibits identical distributions as benign traffic over time. This is an extreme version of concept drift~\cite{lu2018learning} and is better addressed by introducing more fine-grained features and re-collecting the entire training set.

\section{Detecting Encrypted Malicious Traffic}
\label{sec:methodology}

\begin{figure*}
    \centering
    \includegraphics[width=\textwidth]{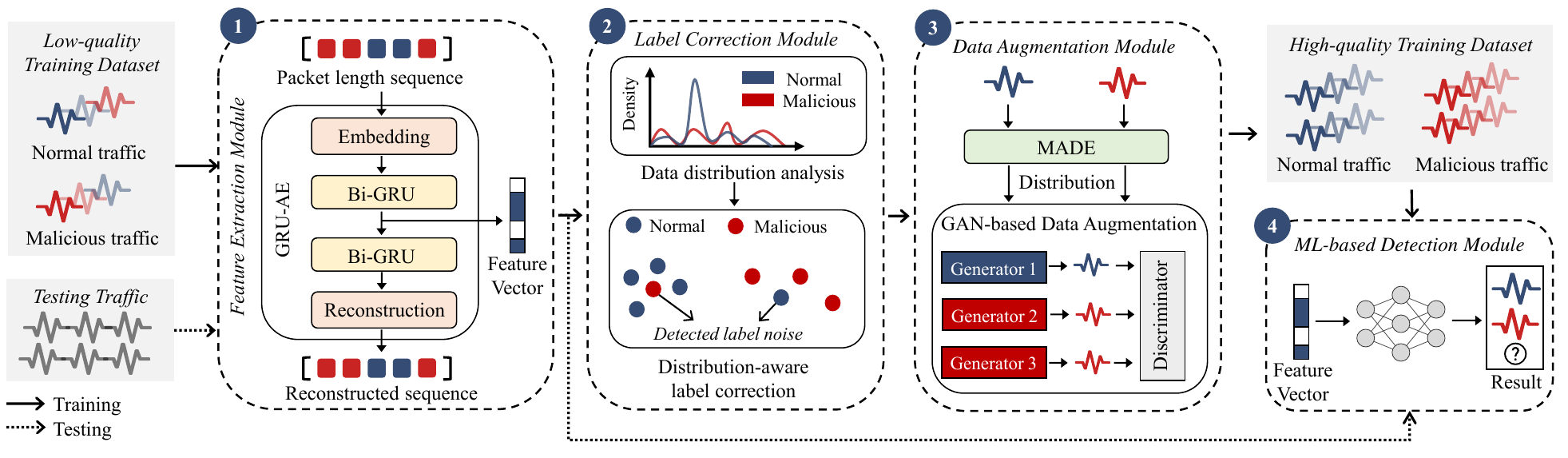}
    \caption{The overview of our robust encrypted malicious traffic detection system \ours.}
    \label{fig:overview}
    \vspace{-0.1in}
\end{figure*}

\subsection{Overview}
\label{subsec:method-overview}

We design our system based on a critical observation that the distribution of benign and malicious data is different. 
First, the normal (\ie benign) traffic is relatively representative. 
\qing{Previous studies on malicious traffic detection regularly use this characteristic of normal traffic when designing their  models~\cite{jan2020odds, mirsky2018kitsune, tang2020zerowall}.}
Second, since network hosts can be infected by a variety of malware, the malicious traffic is relatively manifold. 
Thus, we observe that the distribution of normal data tends to be similar and denser, while that of malicious data (which may be generated by an amount of malware) tends to be more sparse. 
Thereby, given a noisy training set, we can infer the true labels of training samples located in the densest and sparsest parts of its data distribution and use them as a basis for correcting other training samples' labels. Moreover, the distribution of new malicious data may move closer to normal data over time since sophisticated attackers are likely to imitate normal behaviors to avoid detection. 
Thus, we can infer the distribution of new malicious data and then synthesize new training data to improve the performance of ML-based detectors on new unseen testing data.

\ours consists of three main components, including a feature extraction module, a label noise correction module, and a data augmentation module, as shown in Figure \ref{fig:overview}.  The feature extraction module utilizes an improved auto-encoder architecture to convert the raw encrypted traffic into feature vectors that represent fine-grained behaviors while eliminating the negative influence of label noise on feature extraction. Next, the label noise correction module applies an auto-regressive generative model to accurately estimate the distribution of limited yet high-dimensional training data and revise the training samples' labels based on their distribution characteristics. Then, with the label-corrected training data, the data augmentation module infers the possible distribution regions of new malicious data and applies an improved GAN model to synthesize new malicious training data that is located in these target regions. Meanwhile, to prevent the newly generated malicious data from amplifying the degree of data imbalance, this module synthesizes new normal training data that can maintain the decision boundary of ML models. 
Finally, our system utilizes an ML-based detector built upon Co-teaching~\cite{han2018coteaching} and Multilayer Perceptron (MLP) that is trained on the synthetic training data and the label-corrected original training data, to detect the encrypted malicious traffic. \qing{Here, we apply Co-teaching to eliminate impacts of the small amount of the remaining label noises in the training set. Other techniques robust to label noises, \eg decoupling~\cite{malach2017decoupling}, are also applicable.}
In the testing stage, the encrypted traffic is inspected by the built ML-based detector directly after being converted into feature vectors. 

\subsection{Feature Extraction}
\label{subsec:method-feature}
The feature extraction module
converts raw encrypted traffic into feature vectors, which facilitates label noise correction and data augmentation afterward. 
It needs to handle payloads encrypted by various types and versions of encryption  protocols~\cite{feng2021off}. 
The traditional encrypted traffic detection methods~\cite{anderson2016cisco,anderson2017cisco,liu2018mampf} that extract features for specific versions of TLS handshake metadata or message types cannot capture features representing fine-grained behaviors of network traffic. 
\qing{Moreover, existing automatic traffic profiling methods~\cite{aceto2019mimetic, deng2023robust,rimmer2017automated, shen2023subverting, sirinam2018deep} based on supervised models are ill-suited because the non-negligible label noises in the low-quality training set will result in inaccurate feature selection.}

To profile fine-grained behaviors of encrypted network traffic while eliminating the impacts of label noises, our feature extraction module uses an Auto-Encoder architecture (AE) to automatically learn the most representative features of input data in an unsupervised manner and minimize the effects of label noises. It consists of an embedding layer, an encoder, a decoder, and a reconstruction layer. We divide raw encrypted traffic into network flows based on the five-tuple information (\ie source and destination IP addresses, source and destination ports, and the transport layer protocol) and use the sequence of packet lengths of each flow as the input data for feature extraction. Compared with the TLS handshake metadata and message type information, the packet length sequence is more general and can capture the subtle differences among different flows. The encoder and decoder in AE are built based on the bi-directional Gated Recurrent Unit (bi-GRU)~\cite{chung2014empirical}, which allows profiling sequential data in AE. 

We use $l = \left [l_1,l_2,...,l_n\right ]$ to denote the packet length sequence of $n$ packets in a flow, where $l_i$ is the length of the $i$-th packet.
The embedding layer is a learnable matrix $M \in \mathbb{R}^{L \times V}$, where $L$ is the number of individual packet lengths and $V$ is the dimension for embedding. We convert $l$ into an embedding sequence $v =\left [v_1,v_2,...,v_n \right ]$, where $v_i$ is a $V$-dimensional vector by retrieving the $l_i$-th row of $M$. 

The encoder stacks multiple bi-GRUs to learn representative features from the embedding sequence. GRU is one of the most popular neural networks for sequential data. It regards the input sequence as a time series and computes a hidden state at each time step, to represent the key information about the sequence before the time step. The hidden state $h_t$ at time step $t$ is derived from the hidden state $h_{t-1}$ at time step $t-1$ and the input $v_t$ at time step $t$. For simplicity, 
we only emphasize the major computation process formalized as follows: 
\begin{equation}
    h_t = GRU\left ( h_{t-1}, v_t \right ). 
\end{equation}
\noindent where $h_t, h_{t-1} \in \mathbb{R}^H$ and $H$ is the hidden size of GRU layers.

The bi-GRU in the first layer processes the embedding sequence $v$ through two individual GRUs: the forward $\overrightarrow{GRU}$ whose input is from $v_1$ to $v_n$ and the backward $\overleftarrow{GRU}$ whose input is from $v_n$ to $v_1$. Then their hidden states $\overrightarrow{h}_t$ and $\overleftarrow{h}_t$ at time step $t$ ($t\in\left \{ 1,2, \dots, n \right \} $) can be computed as follows:
\begin{equation}
    \overrightarrow{h}_t = \overrightarrow{GRU}\left (\overrightarrow{h}_{t-1}, v_t \right ),
    \overleftarrow{h}_t = \overleftarrow{GRU}\left ( \overleftarrow{h}_{t+1}, v_t \right ).
\end{equation}

\noindent Note the bi-GRU in the $i$-th layer uses the hidden states of the $(i-1)$-th layer as the input data and computes its own hidden states. Since the hidden state at the final time step captures the key information of the whole input sequence, our encoder concatenates the hidden states at the final time steps of all bi-GRU layers as the feature vector $f_{encoder}$ of the length sequence $l$. Assume that the encoder contains $B$ bi-GRU layers, $f_{encoder}$ can be represented as:
\begin{equation}
f_{encoder}=\left ( \overrightarrow{h}_n^{(1)},\overleftarrow{h}_1^{(1)},...,\overrightarrow{h}_n^{(B)},\overleftarrow{h}_1^{(B)} \right ) \in \mathbb{R}^{(2BH)} = \mathbb{R}^d,
\end{equation}
where $d=2BH$ is the dimension of the feature vector.

The decoder applies similarly stacked bi-GRUs to convert the feature vector back to the original embedding sequence. In particular, the bi-GRU in the first layer uses $f_{encoder}$ as the input data for each time step, while the bi-GRUs in other layers work the same way as the ones in the encoder. The output of the decoder is a sequence in length $n$, where the $i$-th element is the hidden state of the bi-GRU in the last layer at time step $i$. Finally, the reconstruction layer restores the packet length sequence $\hat{l}$ from the output of the decoder via a multi-layer perception. In the training stage of our feature extraction module, we compute the reconstruction loss based on $l$ and $\hat{l}$ and apply the stochastic gradient descent optimization algorithm, which allows the encoder to learn accurate representations of encrypted network traffic. 

FS-Net~\cite{liu2019fsnet} is also built upon Auto-Encoder; however, it introduces an additional classification layer and adopts supervised learning via the cross-entropy loss. Thus, it is sensitive to label noises in the training data. In contrast, our system does not require any label information for training and thus enables accurate feature extraction from low-quality data. 

\subsection{Label Noise Correction}
\label{subsec:method-label_correction}

\begin{figure}
    \centering
    \includegraphics[width=\linewidth]{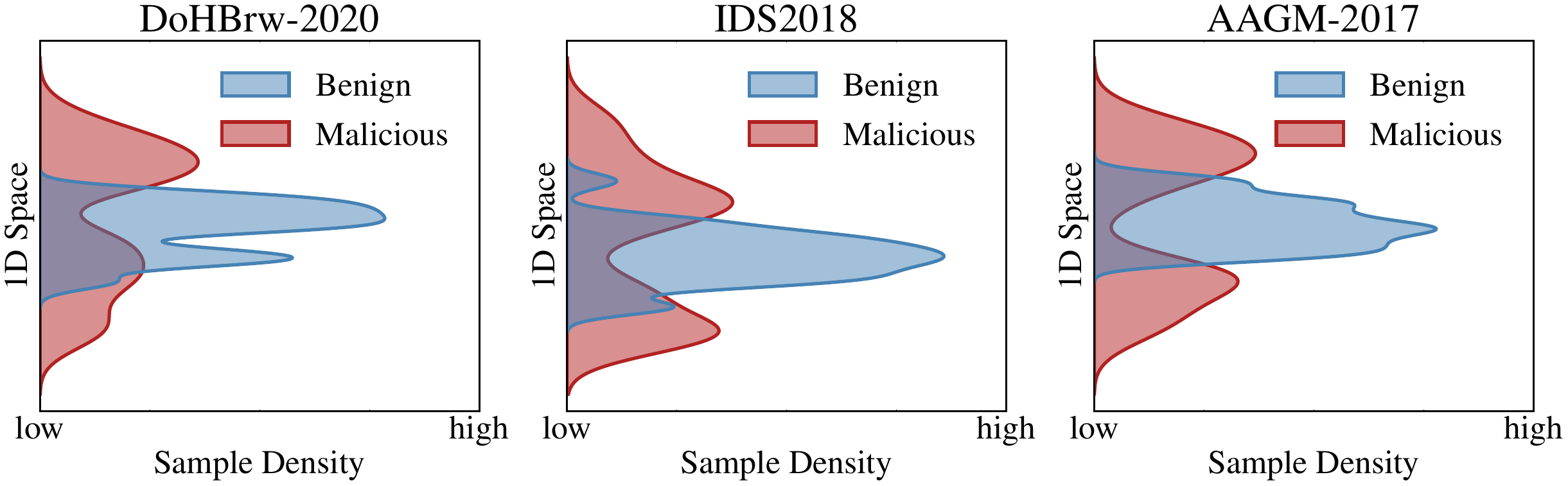}
    \caption{The sample density with normal and malicious data sampled from DoHBrw2020, IDS2018, and AAGM2017 datasets. We reduce the feature dimension of the data to 1 for more intuitive visualizations. The Y-axis represents the locations in 1D space and the X-axis represents the sample density of either type at the specific location in 1D space.}
    \label{fig:density_on_datasets}
    \vspace{-0.15in}
\end{figure}
\begin{figure}
    \centering
    \includegraphics[width=0.8\linewidth]{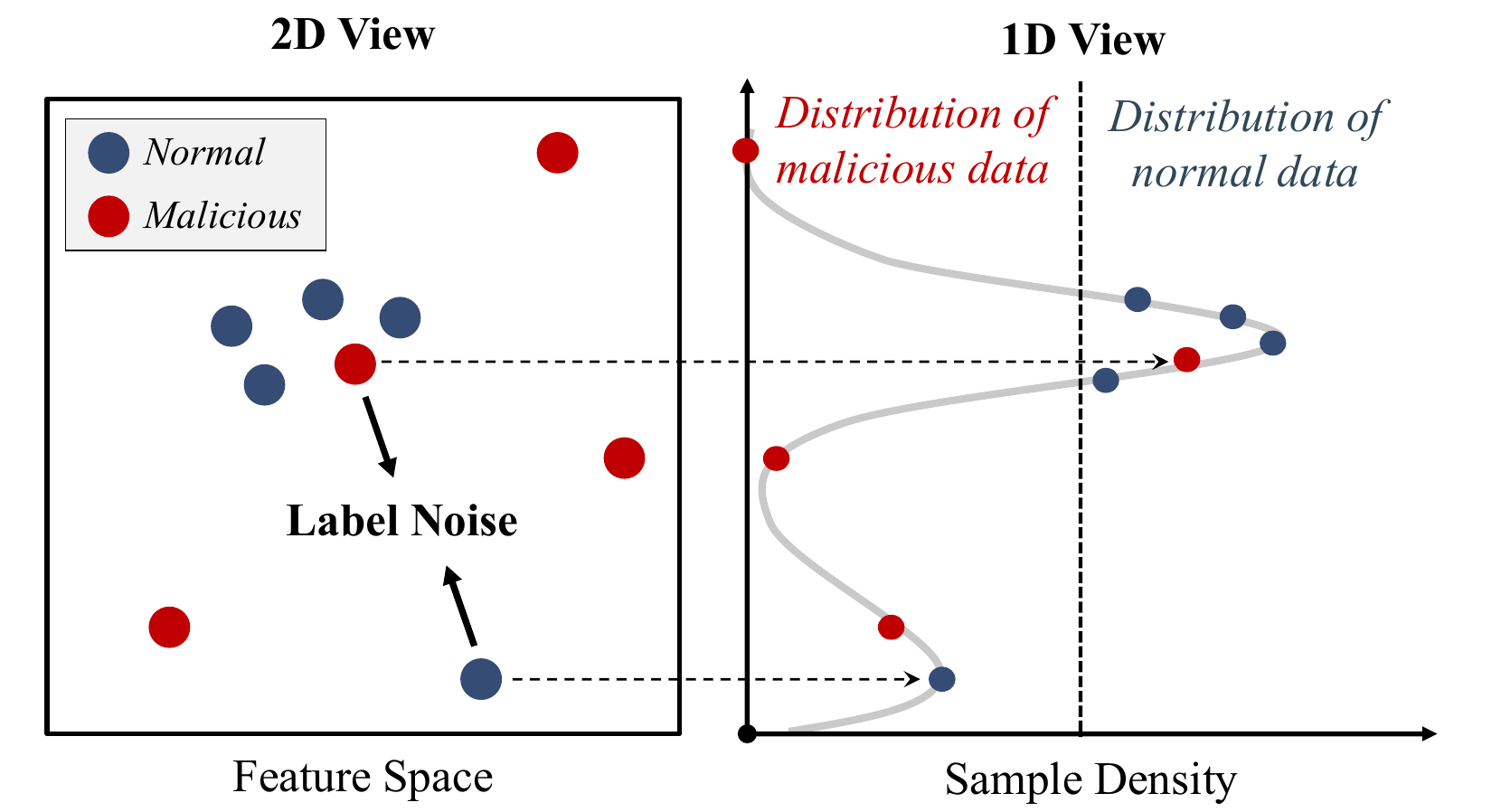}
    \caption{The basic idea of our label noise correction module. The samples with high (\resp low) densities are more likely to be the true normal (\resp malicious) samples.}
    \label{fig:label_correction}
    \vspace{-0.15in}
\end{figure}

The label noise correction module corrects incorrect labels in low-quality training sets by taking advantage of the difference in distribution between normal and malicious data. This approach is based on our observation that normal data tends to have a denser distribution than malicious data.
We validated this observation on three public datasets:  DoHBrw2020~\cite{DoHBrw}, IDS2018~\cite{sharafaldin2018toward} and AAGM2017~\cite{lashkari2017AAGM}. 
For each dataset, we randomly selected 500 benign and 500 malicious network flows. We extracted feature vectors for each flow using our feature extraction module. Next, we used the TSNE~\cite{laurens2008TSNE} algorithm to reduce the dimensionality of the feature vectors to 1. Finally, we estimated the distribution of each type of data using kernel density estimation~\cite{terrell1992variable}. Figure~\ref{fig:density_on_datasets} shows the density of normal data is higher than that of malicious data.

Therefore, we can estimate the distribution of the low-quality training data and then identify the training samples that are located in the densest and sparsest regions of the distribution, as shown in Figure~\ref{fig:label_correction}. These samples can then be labeled as normal and malicious, respectively. 
However, there are a number of challenges to achieving this goal. First, the training data after feature extraction is in the form of high-dimensional feature vectors. To ensure the accuracy of data distribution estimation, it is not appropriate to reduce the feature dimension of the training data. However, traditional statistical methods like kernel density estimation~\cite{terrell1992variable} cannot accurately estimate the distribution of high-dimensional data.
Second, the difference between normal and malicious data distributions may not be significant, especially for the less frequent normal operations. It becomes difficult to infer the true labels of all training samples solely based on their distributions.

We develop our label noise correction module to overcome these issues. First, we estimate the distribution of training data based on the deep generative model~\cite{kingma2014semi}, which can capture the probability distribution of observable variables. Benefiting from the deep architectures, the deep generative model can learn the latent correlation between different features and model high-dimensional data accurately. According to the estimated distribution, we relabel part of the training samples that have the most significant distributions and then utilize them as a basis to infer the true labels of other training data through ensemble learning. 

Specifically, we leverage MADE~\cite{mathieu2015made}, a highly efficient auto-regressive generative model, for distribution estimation. 
MADE models the distribution of input data by learning its probability density function. Assume that $X$ is the input data set for MADE and $x=\left ( x_1,x_2,...,x_d \right ) \in \mathbb{R}^d$ is an sample belonging to $X$. MADE will output a series of 
conditional probability densities $p(x_1)$, $p(x_2|x_1)$, $p(x_3|x_1 x_2)$, \dots, $p(x_d|x_1 x_2 x_3 \dots x_{d-1})$ for $x$. Then, the joint probability density of $x$ can be computed as:
\begin{align}
    \label{MADE_density}
    p(x) &= \prod_{i=1}^d p(x_i|x_1\dots x_{i-1}) \nonumber \\
    &\text{with\quad} p(x_i|x_1\dots x_{i-1}) = p_\mathcal{M}(x_i|\zeta_i).
\end{align}
\noindent where $p_\mathcal{M}$ is a Gaussian mixture function with learnable parameters $\zeta_i$. MADE learns the probability density function of $X$ by maximizing the likelihood of each sample via stochastic gradient descent. 
The probability density of a sample approximately reflects a sample's position in the data distribution, \eg in the densest or the sparsest region, which is helpful for inferring the sample's true label. 

We select the set of training samples having normal labels (denoted as $\widetilde{w}_{train}$) from the original low-quality training set $\widetilde{D}_{train}$ and only use $\widetilde{w}_{train}$ as the input data set for MADE. Since $\widetilde{w}_{train}$ usually contains less true malicious data than $\widetilde{D}_{train}$\footnote{Here, we do not consider datasets with over 50\% data mislabeled (worse than random labeling). Please see Section~\ref{sec:discussion} for discussion on this issue.}, this strategy prevents MADE from assigning high-density values to the true malicious data. It also effectively enlarges the difference between truly normal and malicious data. 
Then, we obtain the probability density values of each training sample in $\widetilde{D}_{train}$ and take the following steps to relabel some samples. First, we select a set $\widetilde{H}_{train}$ from $\widetilde{D}_{train}$, where the samples in $\widetilde{H}_{train}$ have higher probability density values than the ones outside the set. The size of $\widetilde{H}_{train}$ is set to $\alpha \cdot \left | \widetilde{D}_{train} \right | $ and $\alpha \in \left ( 0,1 \right ) $ is a pre-defined parameter. Second, we measure its average Euclidean distance to others for each sample in $\widetilde{H}_{train}$ 
and select half samples with smaller distance values. The selected samples, denoted as $N_s$, are located in the densest regions of the data distribution and are similar to each other. Thereby, we can relabel them to be normal training data with high confidence. Third, for each sample in $\widetilde{D}_{train}-N_s$, we also compute its average Euclidean distance to all samples in $N_s$ and select another set of samples $M_s$ with larger distances. The samples in $M_s$ are located in the sparse regions of the data distribution and are least similar to $N_s$. Thus, we relabel $M_s$ to be malicious training data. Meanwhile, we set the size of $M_s$ the same as $N_s$ to prevent data imbalance in inferring the true labels of other samples.

Finally, with the label-corrected sets $N_s$ and $M_s$, we infer the true labels of other samples in $\widetilde{D}_{train}$ through ensemble learning. Instead of applying a single classifier, we choose ensemble learning since it can improve the classification accuracy under limited training data and is more robust to label noises. We build an ensemble of seven classical machine learning classifiers based on $N_s$ and $M_s$, including Linear Discriminant Analysis, AdaBoost, Random Forest, Logistic Regression, Gaussian Naive Bayes, SVC, and XGBoost, to predict the true label (\ie normal or malicious) of each sample in $\widetilde{D}_{train}-N_s-M_s$. \qing{This ensemble classifier uses the feature vectors extracted by the feature extraction module.} Thus, we can obtain 
the label-corrected training set $D_{train}$, 
and the training samples with corrected labels are denoted as $D_{normal}$ and $D_{malicious}$, respectively. 

\subsection{Data Augmentation}
\label{subsec:method-data_augmentation}

The data augmentation module aims to synthesize new training data that can improve the detection performance on unseen testing data according to the label-corrected original training set $D_{train}$, which is non-trivial due to the following reasons: \first We should ensure the diversity of the new training data. Otherwise, the machine learning model may still overfit the new training data, limiting its generalizability. \second The distribution of the testing data is likely to be inconsistent with the training data because malware is always evolving. As a result, the traditional data oversampling methods~\cite{chawla2002smote, han2005BorderlineSMOTE, he2008adasyn} and the popular GAN-based data synthesis approaches~\cite{frid2018synthetic,goodfellow2014GAN} are not applicable. Specifically, the oversampling methods essentially replicate the original training data, while the vanilla GAN model can only generate new data that follows the distribution of the original training data. 
To address these issues, we predict possible distribution regions of new malicious data and sample from these target regions to synthesize new diverse training data. 

\begin{figure}
    \centering
    \includegraphics[width=0.8\linewidth]{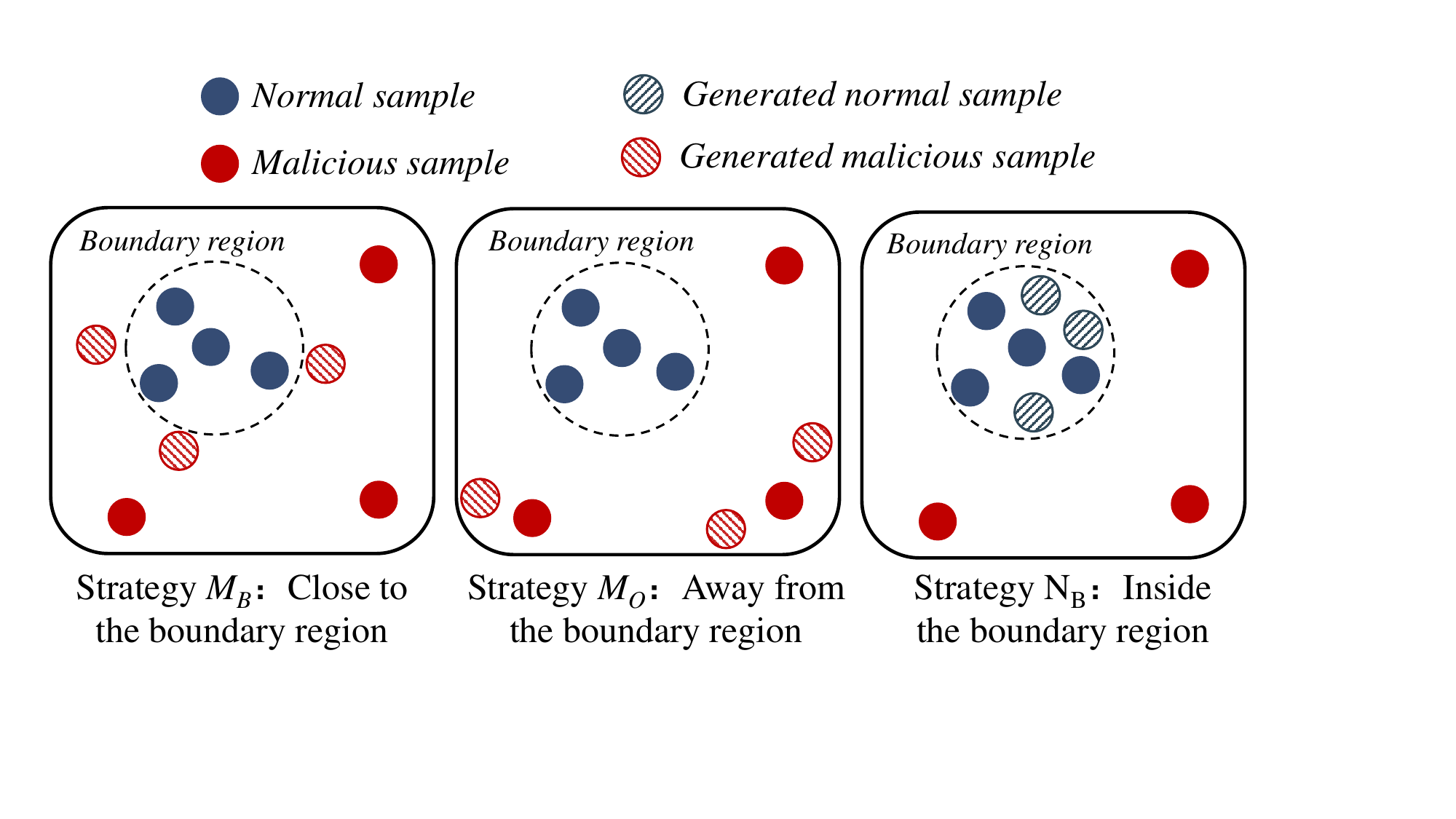}
    \caption{Three data augmentation strategies: $M_B$ and $M_O$ are for augmenting malicious samples and  $N_B$ is for augmenting normal samples.}
    \label{fig:data-augmentation-strategy}
    \vspace{-0.15in}
\end{figure}

Our data augmentation module augments training data by synthesizing both normal and malicious data according to data distributions. We argue that new malicious data may be located in two specific regions following the data distribution. First, considering that sophisticated attackers tend to imitate normal behaviors to evade detection, new malicious data may become more similar to normal data over time. Therefore, in the data distribution, some new malicious data may be located in the boundary region outside the normal data and yet very close to it (denoted as $M_B$). Second, since new attack methods are emerging, \eg Zero-day attacks, new malicious data may be different from both normal and original malicious data, \ie locating in the region outside all original training data (denoted as $M_O$). However, we cannot simply synthesize new malicious training data in these two areas, not 
exacerbating data imbalance. 
For example, when there is far more malicious data than normal data in the region, the decision boundary of the machine learning model will move toward the region of normal data and the model is more likely to predict malicious results, \ie generating more false alarms.  
Thus, our module chooses the boundary region inside the original normal data to guide the generation of normal data (denoted as $N_B$), which is essential for maintaining the decision boundary of the machine learning model. \qing{We visualize these strategies in Figure~\ref{fig:data-augmentation-strategy}.}

Since the probability density can reflect a sample's location in the distribution, we use probability density functions to describe the target regions. In particular, we estimate the data distribution of $D_{normal}$ and $D_{malicious}$ via MADE and use $p_N\left ( \cdot  \right )$ and $p_M\left ( \cdot  \right )$ to denote the corresponding probability density functions, respectively. The target regions corresponding to the three augmentation strategies can be represented as:
\begin{equation}
    \label{Data_augmentation_MN}
    \begin{aligned}
        M_B & = \{ x| p_M\left ( x \right )< \gamma, & \omega_1 & \leq p_N\left ( x \right ) < \omega_2 \}, \\
        M_O & = \{ x| p_M\left ( x \right )< \gamma, & 0 & \leq p_N\left ( x \right )< \omega_1 \}, \\
        N_B & = \{ x| p_M\left ( x \right )< \gamma, & \omega_2 & \leq p_N\left ( x \right ) < \omega_3 \}, 
    \end{aligned}
\end{equation}
\noindent where $ x\in \mathbb{R}^d$, $\omega_1$, $\omega_2$, $\omega_3$, $\gamma$ are pre-defined thresholds to control the sizes and locations of the target regions. Then, we represent the new training data sampled from each target region in the form of distribution, which can be uniformly defined as:
\begin{equation}
    P(x)=
    \begin{cases}
        C & p_M(x) < \gamma,  \theta_1 \leq p_N(x) < \theta_2, \\
        \tau_1 \cdot \frac{1}{p_M(x)} & p_M(x) \geq \gamma, \\
        \tau_2 \cdot p_N(x) & p_M(x) < \gamma, p_N(x) < \theta_1, \\
        \tau_3 \cdot \frac{1}{p_N(x)} & p_M(x) < \gamma, p_N(x) \geq \theta_2, \\
    \end{cases}
    \label{Data_augmentation_P}
\end{equation}
\noindent where $C$ is a constant and $\tau_1, \tau_2, \tau_3$ are regularization terms. When applying this distribution to represent the data sampled from a specific target region, $\theta_1$ and $\theta_2$ will be replaced with the parameters relevant to this region, \eg $\omega_1$ and $\omega_2$ for $M_B$. This distribution means that we perform uniform sampling in the target region to generate new data while reducing the probability of new data being sampled from the areas outside the target region. 

\begin{figure}
    \centering
    \includegraphics[width=0.85\linewidth]{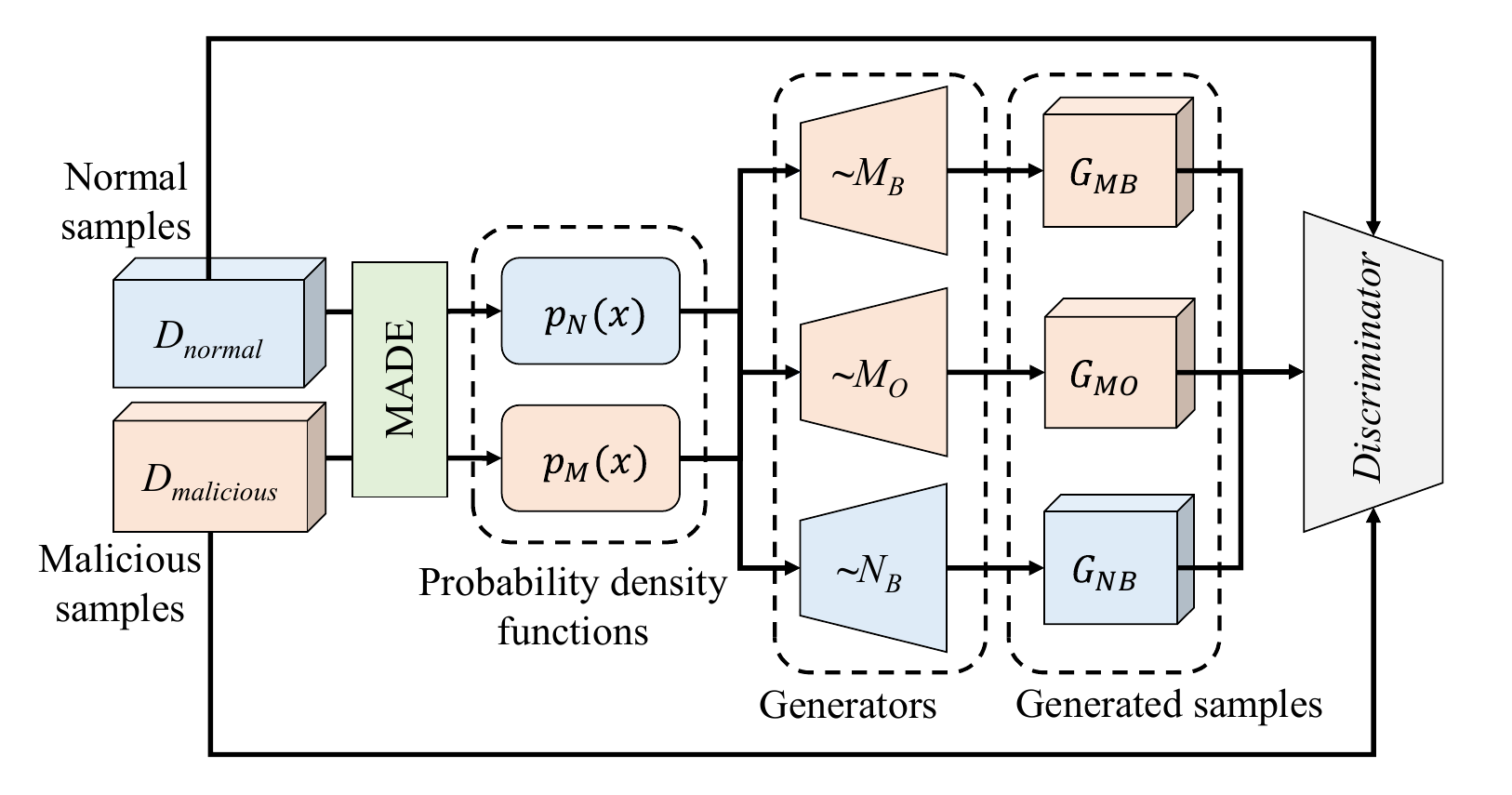}
    \caption{The overview of our data augmentation module. We use three generators to generate samples ($G_{MB}$, $G_{MO}$, $G_{NB}$) based on three strategies ($M_B$, $M_O$, $N_B$), respectively. The \textasciitilde$M_B$ indicates the generator that fits the target distribution region of $M_B$, and the same goes for \textasciitilde$M_O$ and \textasciitilde$N_B$.}
    \label{fig:data-augmentation-model}
    \vspace{-0.15in}
\end{figure}

To synthesize new data that follows the distributions of the three target regions, we develop an improved GAN~\cite{goodfellow2014GAN} model for data augmentation, as shown in Figure~\ref{fig:data-augmentation-model}. 
Unlike the vanilla GAN model approximating the distribution of the original training data, our model utilizes customized loss functions to learn the target data distributions. It consists of three \emph{generator} networks and a \emph{discriminator} network. Each \emph{generator} is responsible for synthesizing new data that follows a target distribution, and the \emph{discriminator} categorizes the newly generated data to perform adversarial training. We use $X_G$ to denote the distribution of the new data generated by one \emph{generator}. The training goal of this \emph{generator} is to minimize the Kullback-Leibler (KL) divergence between $X_G$ and $P(x)$:
\begin{equation}
    \begin{aligned}
        \mathcal{L}_{KL({X_{G}} || P)} = 
            & - \mathcal{H}(X_{G}) \\
            & + \mathbb{E}_{x \in X_{G}, p_M(x) \geq \gamma} \left [ \log p_M(x) \right ] \\
            & - \mathbb{E}_{x \in X_{G}, p_M(x) < \gamma, p_N(x) < \theta_1 } \left [ \log p_N(x) \right ] \\
            & + \mathbb{E}_{x \in X_{G}, p_M(x) < \gamma, p_N(x) \geq \theta_2 } \left [ \log p_N(x) \right ],
        \label{Data_augmentation_KL}
    \end{aligned}
\end{equation}
\noindent where $\mathcal{H}$ is the entropy function that can be approximated by the pull-away term \cite{zhao2017energybased}.
Note that the minimum $p_N\left(x \right)$ threshold of $M_O$ is 0 (shown in Eq~\eqref{Data_augmentation_MN}), which means that all areas in the feature space without normal samples located in the $M_O$ region. This allows us to generate new samples in the area containing no existing samples and makes our data augmentation method superior to the vanilla GAN.

Since network traffic always follows specific protocol specifications, we need to prevent newly synthesized data from deviating from the original training data. Thereby, for one \emph{generator}, we select the original training samples that already locate in the corresponding target region, \ie $X_{in} = \{x | x \in D_{train}, p_M(x) < \gamma, \theta_1 \leq p_N(x) < \theta_2\}$, and minimize their difference with the new data generated by this \emph{Generator}. This goal can be defined as:
\begin{equation}
    \mathcal{L}_{f} = {||} \mathbb{E}_{x \in X_G} \left [ \mathcal{D}_f(x) \right ] 
                        -\mathbb{E}_{x \in X_{in}} \left [ \mathcal{D}_f(x) \right ] {||}_2,
    \label{Data_augmentation_Lf}
\end{equation}
where $\mathbb{E}$ is the expectation function and $\mathcal{D}_f$ is the first layer of the \emph{discriminator}. Then, the complete loss function for one \emph{generator} is as follows:
\begin{equation}
    \mathcal{L}_{G} = \mathcal{L}_{KL({X_{G}} || P)} + \mathcal{L}_{f}.
    \label{Data_augmentation_LG}
\end{equation}

\indent We denote the new data generated by three \emph{generator} networks (\ie \textasciitilde$M_B$, \textasciitilde$M_O$ and \textasciitilde$N_B$) as ${G_{MB}}$, ${G_{MO}}$ and ${G_{NB}}$, respectively.
They are fed into the \emph{discriminator} with $D_{train}$ (including $D_{normal}$ and $D_{malicious}$). 
Similar to the vanilla GAN, our \emph{generator} generates new data that follows target distributions with high diversity. 
Our \emph{discriminator} is designed to classify all data, including the original and synthesized data, into malicious and normal categories. This allows the \emph{generator} to generate new data that can improve the classification performance of machine learning models. In particular, we define $\mathcal{D}(x)$ as the prediction of $x$. $\mathcal{D}(x) \in [0, 1]$, which will be higher if $x$ is more likely malicious. The loss function for our \emph{discriminator} is defined as:
\begin{equation}
    \label{Data_augmentation_LD}
    \begin{aligned}
        \mathcal{L}_D & = \mathbb{E}_{x \in D_{normal}} \left [ log(\mathcal{D}(x)) \right ] + \mathbb{E}_{x \in D_{malicious}} \left [ log(1 - \mathcal{D}(x)) \right ] \\
                      & + \mathbb{E}_{x \in G_{MB}} \left [ log(1 - \mathcal{D}(x)) \right ] + \mathbb{E}_{x \in G_{MO}} \left [ log(1 - \mathcal{D}(x)) \right ] \\
                      & + \mathbb{E}_{x \in G_{NB}} \left [ log(\mathcal{D}(x)) \right ]. \\
    \end{aligned}
\end{equation}

Our GAN model is trained with the label-corrected original data $D_{train}$ by adversarial training. When the model converges, the new data synthesized by the \emph{generator} networks will be regarded as the new training data. Besides, to avoid the model collapse problem~\cite{bau2019seeing} and further increase the diversity of synthesized data, our data augmentation module trains several independent GAN models and combines all their synthesized data as the new training data. Then, we can fully train a machine learning model based on new training data and $D_{train}$ to perform accurate detection. 

Our method addresses the limitations in the ODDS-based data augmentation method~\cite{jan2020odds} that generates new malicious data outside the distribution of original training data. Our model synthesizes new malicious data located in the boundary region outside the normal data to effectively detect new attack methods that mimic normal behaviors. Also, our model synthesizes new normal samples that maintain the decision boundary of machine learning models, reducing false alarms. Furthermore, we train multiple GAN models independently to avoid the model collapse problem, ensuring the diversity of new data.

\section{Evaluation}
\label{sec:evaluation}

We evaluate the effectiveness of our detection system using public and real-world datasets. We also compare the performance of our work with representative encrypted malicious traffic detection methods.

\subsection{Experimental Setup}
\label{subsec:set-up}

\begin{table}[t]
\centering
\small
\renewcommand{\arraystretch}{1.1} 
\caption{The number of normal and malicious encrypted network flows in the public datasets. The flows generated on the first day of the dataset are regarded as T1 and the rest is as T2. We randomly select a small amount of training data from T1 in each experiment and the whole T2 is used for testing.}
\label{tab:datasets}
\begin{tabular}{c|cc|cc}
\toprule

\multirow{2}{*}{Dataset}
& \multicolumn{2}{c}{T1} & \multicolumn{2}{c}{T2} \\ \cmidrule(lr){2-3}  \cmidrule(lr){4-5} 
 & Normal & Malicious  
 & Normal & Malicious \\ \midrule 
DoHBrw & 
4381 & 10,298 & 304,313 & 138,449\\
IDS & 
430,974 & 6,446 & 1,502,583 & 47,186\\ \bottomrule
\end{tabular}
\end{table}

\noindent \textbf{Datasets.} We use the following public datasets for evaluation:
\begin{itemize}[itemsep=0em,align=parleft,left=0pt..1em]
    \item \textbf{CIRA-CIC-DoHBrw-2020 (DoHBrw)} \cite{DoHBrw} includes normal and malicious DNS-over-HTTPS (DoH) encrypted traffic. The normal traffic is generated by querying benign DNS servers, \eg Cloudflare and Google, using the DoH protocol. The malicious DoH traffic is generated by three different DNS tunneling tools including dns2tcp, DNSCat2, and Iodine. It encapsulates the DNS tunnel data communicating with malicious DNS servers. The dataset is collected at a location between clients and the gateway, and all data is HTTPS traffic. 
    
    \item \textbf{CSE-CIC-IDS2018 (IDS)} \cite{sharafaldin2018toward} is a widely-used intrusion detection dataset that records the network traffic generated by hundreds of hosts in an internal network and includes seven attack scenarios. We extract the encrypted traffic from this dataset as our evaluation data. However, it has only a small amount of malware's encrypted traffic. Thus, we use the CIC-InvesAndMal2019~\cite{taheri2019extensible} dataset, which contains the malicious encrypted traffic generated by 426 malware samples in six malware types (\ie Adware, Botnet, PremiumSMS, Ransomware, Scareware, and SMS) to supplement our evaluation data. Besides, to ensure the normal encrypted traffic used for evaluation is correctly labeled, we filter it based on the Alexa Top list and remove the traffic generated during the attacking time. 
    
\end{itemize}

Table \ref{tab:datasets} shows each dataset is divided into two sets according to the timestamps of traffic. We put all the traffic generated on the first day of the dataset into set T1 and the remaining data into set T2. In each experiment, we randomly select a small amount of traffic from T1 as the training set and utilize the entire T2 as the testing set. 

\noindent \textbf{Baselines.}
We choose several state-of-art malware detection methods and two representative machine learning methods that can deal with low-quality training data, to design the baselines:

\begin{itemize}[itemsep=0em,align=parleft,left=0pt..1em]
    \item \textbf{Malicious Encrypted Traffic Detection Methods.} We select two state-of-the-art flow-level malicious traffic detection methods: \textsf{ETA}~\cite{anderson2017cisco} and \textsf{FS-Net (FS)}~\cite{liu2019fsnet}. \textsf{ETA} utilizes the features relevant to packet length information and TLS handshake metadata and performs detection based on the random forest algorithm. \textsf{FS} is an end-to-end deep learning model that profiles encrypted traffic via time-serial features and performs well in multiple encrypted traffic classification tasks. 
    More crucially, except for a labeled training set, these two methods do not need extra data, \eg a large-scale unlabeled training set, such that we can make fair comparisons with our system. 

    \item \textbf{Robust Malware Detection Methods.} We select two malware detection methods robust to low-quality training data: \textsf{Differential Training (DT)}~\cite{xu2021differential} and \textsf{ODDS}~\cite{jan2020odds}. \qing{These two methods handle label noises and insufficient training data, respectively.} In particular, \textsf{DT} is a generic framework to reduce label noises for Android malware detection. It characterizes malware samples via host-level features, \eg API calls, identifies mislabeled malware training samples by applying several outlier detectors, revises their labels, and then builds ML-based malware detectors based on the label-corrected training set. Besides, \textsf{ODDS} targets the malicious network request launched by bot malware. It synthesizes new malicious data via GAN to improve the performance of the ML-based detector with limited training data.
    \item \textbf{Robust Machine Learning Methods.} We also choose two representation machine learning methods that can handle low-quality training data: \textsf{SMOTE}~\cite{chawla2002smote} and \textsf{Co-teaching (Co)}~\cite{han2018coteaching}. \qing{These methods can handle low-quality training data in different domains.} Specifically, \textsf{SMOTE} is a data augmentation technique that is widely used to address the problem of insufficient training data. SMOTE works by generating synthetic data points that are similar to the minority class. 
    \textsf{Co-teaching} is a general machine learning framework that is robust to label noise. \textsf{Co-teaching} works by training two identical models in parallel. The models are then compared to each other, and any data points that are misclassified by both models are removed from the training set. This helps to improve the accuracy of the models by removing noisy data points.
        
\end{itemize}

Among the above methods, only \textsf{ETA} and \textsf{FS} are designed for encrypted malicious traffic detection. 
For comprehensive comparisons, we produce more baselines by enhancing existing methods with \textsf{ETA} and \textsf{FS}. Specifically, since \textsf{DT} is a general framework for malware detection, we replace its original host-level features and malware detector with the time-serial features used by \textsf{FS} and the \textsf{FS}-based detector, respectively, to create the baseline \textsf{DT+FS}. Also, we enhance \textsf{ODDS} with \textsf{FS} in a similar way to obtain the baseline \textsf{ODDS+FS}. Moreover, we design a more powerful baseline \textsf{DT+ODDS+FS} based on \textsf{DT}, \textsf{ODDS}, and \textsf{FS}. For a low-quality training set, \textsf{DT+ODDS+FS} corrects the label noise by utilizing \textsf{DT} and synthesizes new training data through \textsf{ODDS}, and then it trains \textsf{FS} based on the quality-improved training set. We also generate another baseline \textsf{DT+ODDS+ETA} by integrating \textsf{DT}, \textsf{ODDS}, and \textsf{ETA} in a similar way. Besides, the baselines \textsf{SMOTE+FS} and \textsf{Co+FS} utilize \textsf{SMOTE} and \textsf{Co} to improve the quality of the training set, respectively, and then train \textsf{FS} to perform detection. 

\begin{table}[ht]
    \centering
    \renewcommand{\arraystretch}{1.1} 
    \caption{The Parameter Settings of \ours. In Data Augmentation, $\gamma=0.05$ means $\gamma$ is the 5th percentile of the set $\left \{p_M(x) | x \in M \right \}$. $\omega_1=0.1$, $\omega_2=0.2$ and $\omega_3=0.3$ represent the 10th, 20th and 30th percentile of the set $\left \{p_N(x) | x \in N \right \}$.}
    \label{tab:parameters} 
    \resizebox{0.49\textwidth}{!}{
        \begin{tabular}{c|c|c|c}
            \toprule
            Module & Para. & Value & Description \\
            \midrule
            \multirow{5}{*}{\makecell[c]{Feature \\ Extraction}} & $n$ & 50 & Number of used head packets\\
             & $V$ & 32 & Embedding size of GRU-AE \\
             & $H$ & 8 & Hidden size of each GRU layer \\
             & $B$ & 2 & Number of GRU layers \\
             & $d$ & 32 & Dimension of feature vector ($d = 2BH$) \\
            \midrule
            \multirow{2}{*}{\makecell[c]{Label \\ Correction}} & \multirow{2}{*}{$\alpha$} & \multirow{2}{*}{0.5} & \multirow{2}{*}{Filtering proportion at the first step}\\ 
            & & & \\
            \midrule
            \multirow{5}{*}{\makecell[c]{Data \\ Augmentation}} 
            & $\gamma$   & 0.05  & \multirow{4}{*}{\makecell{Thresholds to control the size and location\\ of target regions for synthesizing new training data}}\\
            & $\omega_1$ & 0.1 & \\
            & $\omega_2$ & 0.2 & \\
            & $\omega_3$ & 0.3 & \\
            & $\eta$ & 5 & Number of independent GAN models\\
            \bottomrule
        \end{tabular}
    }
\end{table}

\noindent \textbf{Implementation\footnote{The source code of \textsf{RAPIER} is publicly available at https://github.com/XXnormal/RAPIER.}.} 
We implement our detection system and all baselines by using Python 3.8.5, and the used libraries include NumPy 1.21.2, Pytorch 1.7.1, Tensorflow 2.10.0, and Cuda 11.7. We run these models on a Linux server (5.4.0-126-generic) with Intel(R) Xeon(R) E5-2650 v4 CPU and NVIDIA GeForce RTX 2080Ti GPU. We list all the parameters used by our system in Table~\ref{tab:parameters}. In particular, we set $n=50$ because the head packets of each flow usually contain rich information, \eg the handshake process between the client and server sides, which is beneficial for distinguishing malicious and benign encrypted traffic. We set $d=32$ to make a balance between the complexity of the feature extraction module and its capability of capturing fine-grained behaviors of network traffic.
Then, we apply Grid Search to find the appropriate values for $\alpha$, $\gamma$, $\omega_1$, $\omega_2$, and $\omega_3$. In particular, we set 0.5 for $\alpha$, and the values of $\gamma$, $\omega_1$, $\omega_2$, and $\omega_3$ are 0.05, 0.1, 0.2, and 0.3, respectively. Besides, we set $\eta=5$ for synthesizing more diverse training data. In \S~\ref{sec:evaluation:parameter}, we perform additional experiments to evaluate the performance of our system under different parameter values.

\noindent \textbf{Metrics.} We utilize three common metrics to evaluate the performance of our detection system, including precision, recall, and F1 score. We regard malicious and normal encrypted network flows as positive and negative samples, respectively. With the detection result and the ground truth of one testing set, we can calculate the number of true positive samples (TP), the number of false positive samples (FP), and the number of false negative samples (FN). Then, the three metrics can be computed as follows: $\textsf{precision} = \frac{TP}{TP+FP}$, $\textsf{recall} = \frac{TP}{TP+FN}$, and $\textsf{F1 Score}=\frac{2 \times \textsf{precision} \times \textsf{recall}}{\textsf{precision} + \textsf{recall}}$.

\begin{figure*}[t]
  \centering
  \includegraphics[width=0.85\textwidth]{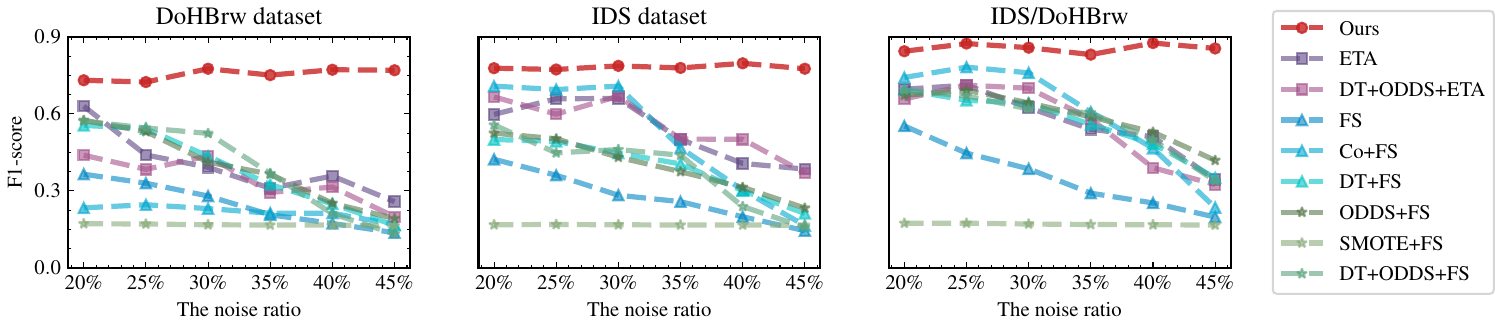}
  \caption{The F1 score of all methods under different noise ratios on three evaluation datasets.}
  \label{fig:noise_ratio}
  \vspace{-0.05in}
\end{figure*}

\begin{table*}[ht]
\centering
\small
\renewcommand{\arraystretch}{1} 
\caption{The F1 score of all methods under different training sizes with the noise ratio of 30\%. The F1 score is shown in the form of "Avg $\pm$ Std", where Avg is the average F1 score and Std is the standard deviation.}
\label{tab:training_size}
\resizebox{\textwidth}{!}{
    \begin{tabular}{c|c|ccccccccc}
    \toprule
    \multicolumn{2}{c|}{Method} & \textsf{FS} & \textsf{Co+FS} & \textsf{DT+FS} & \textsf{ODDS+FS} & \textsf{SMOTE+FS} & \textsf{DT+ODDS+FS} & \textsf{ETA} & \textsf{DT+ODDS+ETA} & \textsf{Ours} \\
    \midrule
    \multirow{3}{*}{\makecell[c]{Training Size on \\ DoHBrw}} & 250 & .18 $\pm$ .02 & .21 $\pm$ .06 & .41 $\pm$ .02 & .42 $\pm$ .05 & .17 $\pm$ .00 & .34 $\pm$ .10 & .56 $\pm$ .27 & .44 $\pm$ .21 & \textbf{.71} $\pm$ .02 \\
     & 500 & .28 $\pm$ .04 & .23 $\pm$ .06 & .44 $\pm$ .03 & .42 $\pm$ .07 & .17 $\pm$ .00 & .52 $\pm$ .04 & .39 $\pm$ .24 & .44 $\pm$ .23 & \textbf{.78} $\pm$ .02 \\
     & 1000 & .30 $\pm$ .03 & .27 $\pm$ .05 & .54 $\pm$ .02 & .57 $\pm$ .06 & .17 $\pm$ .00 & .58 $\pm$ .05 & .40 $\pm$ .23 & .29 $\pm$ .07 & \textbf{.78} $\pm$ .02 \\
     \midrule
     \multirow{3}{*}{\makecell[c]{Training Size on \\ IDS}} & 250 & .26 $\pm$ .05 & .66 $\pm$ .19 & .34 $\pm$ .01 & .46 $\pm$ .04 & .17 $\pm$ .00 & .38 $\pm$ .03 & .57 $\pm$ .10 & .53 $\pm$ .15 & \textbf{.75} $\pm$ .04 \\
     & 500 & .28 $\pm$ .04 & .71 $\pm$ .06 & .45 $\pm$ .04 & .43 $\pm$ .06 & .17 $\pm$ .00 & .46 $\pm$ .08 & .66 $\pm$ .27 & .67 $\pm$ .24 & \textbf{.79} $\pm$ .02 \\
     & 1000 & .35 $\pm$ .01 & .73 $\pm$ .08 & .51 $\pm$ .03 & .51 $\pm$ .08 & .17 $\pm$ .00 & .54 $\pm$ .05 & .66 $\pm$ .27 & .69 $\pm$ .23 & \textbf{.77} $\pm$ .02 \\
     \midrule
     \multirow{3}{*}{\makecell[c]{Training Size on \\ IDS/DoHBrw}} & 250 & .31 $\pm$ .13 & .66 $\pm$ .19 & .51 $\pm$ .10 & .58 $\pm$ .12 & .17 $\pm$ .01 & .53 $\pm$ .11 & .59 $\pm$ .12 & .57 $\pm$ .15 & \textbf{.82} $\pm$ .09 \\
     & 500 & .39 $\pm$ .14 & .76 $\pm$ .06 & .64 $\pm$ .13 & .64 $\pm$ .15 & .17 $\pm$ .00 & .62 $\pm$ .14 & .62 $\pm$ .27 & .70 $\pm$ .27 & \textbf{.86} $\pm$ .05 \\
     & 1000 & .38 $\pm$ .13 & .73 $\pm$ .08 & .68 $\pm$ .09 & .69 $\pm$ .13 & .17 $\pm$ .00 & .72 $\pm$ .13 & .69 $\pm$ .24 & .73 $\pm$ .19 & \textbf{.81} $\pm$ .07 \\
    \bottomrule
    \end{tabular}
}
\end{table*}

\subsection{Overall Detection Performance}
\label{subsec:detect-performance-results}

We first evaluate the detection performance of our system and all baselines on the two public datasets, \ie DoHBrw and IDS. Further, to evaluate \ours an extreme scenario of malware evolution, we create a new dataset IDS/DoHBrw. In particular, we combine all the malicious samples in the DoHBrw dataset with the whole IDS dataset and split it into a new training set and a new testing set according to the timestamp of each network flow. Note that, since the generation time of the DoHBrw dataset (2020) is later than that of the IDS dataset (2018-2019), all the malicious samples of the DoHBrw dataset are put in the new testing set, while the new training set is the same as the original training set of IDS. This setting simulates an extreme case that the new malware samples evolve to be irrelevant to the training data. \qing{All packets in the DoHBrw and IDS datasets are encapsulated by the TLS protocol. Our framework extracts features from each TSL flow. Thus, the features generated from these datasets are the same.}

We set three training sizes, 250, 500, and 1000. Each size represents the number of malicious and normal training samples randomly selected from T1 in each experiment. This random selection strategy ensures the diversity of the training samples.
More importantly, for fair comparisons, we follow the label noise ratio settings used in \textsf{Co}~\cite{han2018coteaching} and \textsf{DT}~\cite{xu2021differential}. In particular, the label noise ratio is defined as the proportion of mislabeled samples in all training samples and there are six different noise ratios (20\%, 25\%, 30\%, 35\%, 40\%, and 45\%) for evaluations.

For each training size value $s$ and label noise ratio value, we create 10 independent low-quality training sets. Each training set consists of $s$ malicious and $s$ normal training samples randomly selected from T1.
The labels of the training samples are then symmetrically flipped according to the label noise ratio. For example, when the label noise ratio is 20\%, we change the label of 20\% malicious training samples to normal and 20\% normal training samples to malicious.
Symmetric label flipping is used in \textsf{Co}~\cite{han2018coteaching} and \textsf{DT}~\cite{xu2021differential}. And we also use this label flipping method in \S~\ref{sec:module-experiments},~\ref{sec:evaluation:parameter} and~\ref{subsec:real-world_performance}. Then, for each low-quality training set, we randomly sample five individual testing subsets from T2, where the size of each one is half of T2. \qing{We set the ratio of malicious to normal samples in each testing subset to 1:10, as recommended by Pendlebury et al.~\cite{pendlebury2019tesseract} in their large-scale malware measurement study.}
Given a low-quality training set, we train \ours and all baselines based on it, evaluate their detection performance on the corresponding five testing subsets, and compute the average performance of each method. The performance of each method under a specific combination of training size and noise ratio is measured by the method's average result on the corresponding ten low-quality training sets.

\noindent\textbf{Impact of Noise Ratio.}
We show the F1 score of all methods under different noise ratios in Figure~\ref{fig:noise_ratio} and the training size is set to 500 (\ie 500 normal and 500 malicious training samples, 1000 samples in total). It can be seen that our system outperforms all baselines with significant margins in all settings. For instance, when the noise ratio is as high as 45\%, our system still achieves the best F1 scores of 0.770, 0.776, and 0.855 on the DoHBrw, IDS, and IDS/DoHBrw datasets, respectively, showing average improvements of 352.6\%, 284.3\%, and 214.9\% over the baselines. Note that, the average improvement is the mean value of the improvement of our system to each baseline.
Besides, we observe that the varying noise ratio only slightly affects our detection performance. In particular, when the noise ratio increases from 20\% to 45\%, the F1 score fluctuations of our system on the DoHBrw, IDS, and IDS/DoHBrw datasets are only about 0.052, 0.024, and 0.045, respectively. By contrast, the average F1 score fluctuations of all baselines except \textsf{SMOTE} are above 0.3. The F1 score of \textsf{SMOTE} keeps lower than 0.2 such that we do not 
compute its performance fluctuations.
These results indicate that when facing highly insufficient training data, our system is much more robust to label noise than all baselines. This is because our system effectively corrects the incorrectly labeled samples based on their distinctive distribution characteristics, which is less relevant with the number of correctly labeled samples.
In specific, the number of label noises after label correction is stable as the origin noise ratio changes, and the detailed performances and explanation of the label-correction module are discussed in \S~\ref{sec:module-experiments}.

\noindent\textbf{Impact of Training Size.}
We show the F1 score of all methods under different training sizes in Table~\ref{tab:training_size}. The noise ratio is set to 30\%.
Our system outperforms baselines in all settings. When the training size is as small as 250, our system achieves the F1 scores of 0.705, 0.751, and 0.817 on the DoHBrw, IDS, and IDS/DoHBrw datasets, respectively, with average improvements of 148.5\%, 113.0\%, and 101.1\% to all baselines. It shows that even when plenty of training samples are incorrectly labeled, our system can synthesize new useful training data for improving detecting performance. In contrast, the baselines that solely utilize existing data augmentation methods, \eg \textsf{ODDS+FS} and \textsf{SMOTE+FS}, cannot deal with the negative impacts of label noise. Moreover, our system remains stable under different training sizes. The F1 score fluctuations of our system on the DoHBrw, IDS, and IDS/DoHBrw datasets are 0.077, 0.036, and 0.053, respectively. It shows that our data augmentation strategies are applicable to different low-quality training sets. In contrast, the baselines' performance shows significant degradations as the training size varies, \eg the F1 scores of \textsf{DT+ODDS+FS} are reduced by 0.24, 0.16, and 0.19 on the three datasets when training size decreases from 1000 to 250. We also notice that all methods can achieve higher detection performance on the IDS/DoHBrw dataset than on the IDS dataset since the malicious samples in the DoHBrw dataset are less similar to the normal samples in the IDS dataset. In this case, our system still outperforms all baselines by a significant margin. This is because the new training data synthesized by our system is more representative of the malware samples in the DoHBrw dataset, allowing our system to learn more about the characteristics of malicious traffic. 

We supplement the detection performance of all methods with each training size and noise ratio setting in Appendix~\ref{app:all_performance}. Our system achieves the best performance in most cases. The reason behind this is that our system accurately corrects label noises within low-quality training data and synthesizes new training data that can represent unseen malware samples.

\noindent\textbf{Impact of smaller malicious-to-normal ratios.} \qing{We further measure the performance of our framework with smaller ratios of malicious to normal samples in the testing sets. We select three ratios: 1:20,1:30, and 1:50. The training size is 500 and the noise ratio is 30\%. The results are shown in Table~\ref{tab:lower_ratio_malicious}. We can see that when the ratio increases from 1:10 to 1:30, the F1 scores of our framework only decrease approximately by 6\% and 8\% on the IDS and IDS/DoHBrw datasets, respectively. Even when the ratio reaches 1:50, our framework still achieves about 0.52, 0.68, and 0.75 F1 scores on the DoHBrw, IDS, and IDS/DoHBrw datasets, respectively. This demonstrates that our framework can effectively detect encrypted malicious traffic in real-world networks where malicious traffic is rare. Additionally, the F1 score decreases due to low precision scores. For example, the precision on the DoHBrw dataset decreases from 0.68 to 0.49 when the ratio rises from 1:10 to 1:30, indicating that more false alarms are reported. We will discuss this issue in \S~\ref{sec:discussion}.}
\begin{table}[ht]
\centering
\small
\renewcommand{\arraystretch}{1.1} 
\caption{\qing{The F1 scores of \ours with smaller ratios of malicious to normal samples. The training size is 500 and the noise ratio is 30\%.}}
\label{tab:lower_ratio_malicious}
\resizebox{.4\textwidth}{!}{
\begin{tabular}{c|c|ccc}
\toprule
\multicolumn{2}{c|}{Dataset} & DoHBrw & IDS & IDS/DoHBrw \\ \midrule

\multirow{4}{*}{\makecell{ {Malicious Ratio} \\ {(malicious : normal)} }} 
 & 1:10 & .78 $\pm$ .02 & .79 $\pm$ .02 & .86 $\pm$ .05 \\
 & 1:20 & .71 $\pm$ .02 & .75 $\pm$ .04 & .82 $\pm$ .04 \\
 & 1:30 & .63 $\pm$ .03 & .74 $\pm$ .05 & .79 $\pm$ .05 \\
 & 1:50 & .52 $\pm$ .05 & .68 $\pm$ .10 & .75 $\pm$ .03 \\
\bottomrule

\end{tabular}}
\end{table}

\subsection{Various Settings of Label Noises}
\label{subsec:realistic_setting}

\begin{figure}[t]
  \centering
  \includegraphics[width=\linewidth]{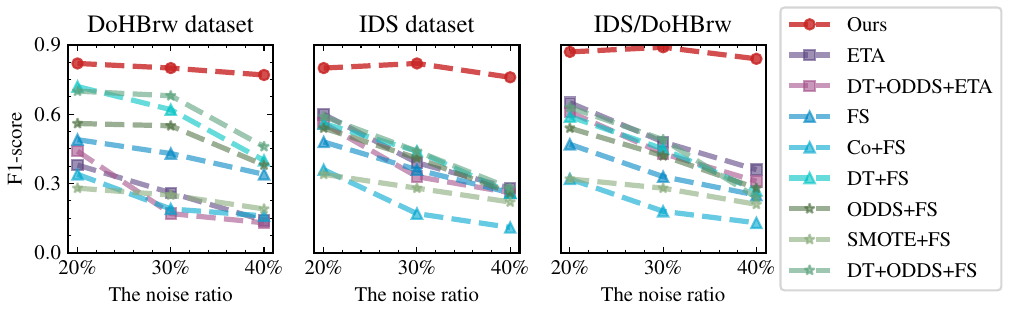}
  \caption{The F1 scores under realistic label noise settings.}
  \label{fig:noise_ratio_real}
\vspace{-0.15in}
\end{figure}

To evaluate the performance of our system under more realistic label noise settings, we create label noise based on two common scenarios where malicious and normal encrypted traffic can be mislabeled. First,  unseen domain names: Since network administrators usually identify normal traffic based on domain names~\cite{antonakakis2011detectDNS}, they may regard an unseen domain name as malicious. Second, absent threat intelligence: When the threat intelligence about certain types of malware is absent, network administrators may not be able to identify the malware's malicious traffic, \ie mislabel the traffic as normal.

For the T1 set of each dataset, we change the label of the normal network flows to malicious if their domain names are not in the Alexa-Top-1m list. Also, we change the label of the malicious network flows to normal if they belong to a chosen malware type (\ie dns2tcp, DNSCat2, Iodine for the DoHbrw dataset and Adware, Botnet, PremiumSMS, Ransomware, Scareware, SMS for the IDS dataset). These network flows with modified labels are considered label noises.
Given a training size and a noise ratio, we choose a malware type to generate the label noises and then create 10 independent low-quality training sets from T1. Each set consists of the label noises and the correctly labeled network flows randomly sampled from T1 according to the given training size and noise ratio. We then train our system and all baselines on each low-quality training set. We create the testing sets using the same method as in \S~\ref{subsec:detect-performance-results} and evaluate each method's performance.

We set the training size to 500 and the noise ratio to 20\%, 30\%, and 40\%. For each training size and noise ratio setting, we choose one malware type at a time and calculate the average detection performance of each method under all malware types. As shown in Figure~\ref{fig:noise_ratio_real}, our system outperforms all baseline models. Specifically, our system achieves average F1 scores (among all settings) of 0.797, 0.800, and 0.867 on the DoHBrw, IDS, and IDS/DoHBrw datasets, respectively, achieving average improvements of 166.5\%, 154.6\%, and 165.2\% over all baselines.
This indicates that our system can perform robust encrypted malicious network detection when trained with the low-quality training data generated in realistic circumstances.
We notice that most baselines' performance decreases compared to their results in \S~\ref{subsec:detect-performance-results} (when samples are randomly selected to be label noises). For instance, the F1 score of \textsf{ETA} declines 30\% on average. In contrast, our method maintains similar detection accuracy, where the F1 score fluctuations are less than 0.03.

\subsection{Evaluating Individual Components}
\label{sec:module-experiments}

\noindent \textbf{Feature Extraction.} We first compare our feature extraction module with two representative feature extraction methods \textsf{ETA}~\cite{anderson2016cisco} and \textsf{CICFlowMeter} \cite{habibi2017CICFlowMeter}. We replace our feature extraction module with these two methods while keeping our label noise correction and data augmentation modules.

The F1 score reduces to 0 when we use these two new feature extraction methods in all settings and 
we inspect the results of two modules.
\textsf{CICFlowMeter} has recall scores of 0, meaning that it is unable to identify encrypted malicious traffic. This is likely because the coarse-grained statistical features extracted by CICFlowMeter are not able to distinguish between encrypted malicious and normal traffic.
Besides, \textsf{ETA} extracts feature vectors of much higher dimensions (386) than our system (32). This makes it difficult for the deep generative model (MADE) to estimate the distribution of such high-dimensional data. As a result, the remaining components of our system depending on the data distribution cannot function properly.
In contrast, we extract more important sequential features that represent fine-grained behaviors, where the extracted feature vectors are of lower dimensions, facilitating accurate distribution estimation performed by the MADE model.

\begin{figure}[t]
  \centering
  \includegraphics[width=0.85\linewidth]{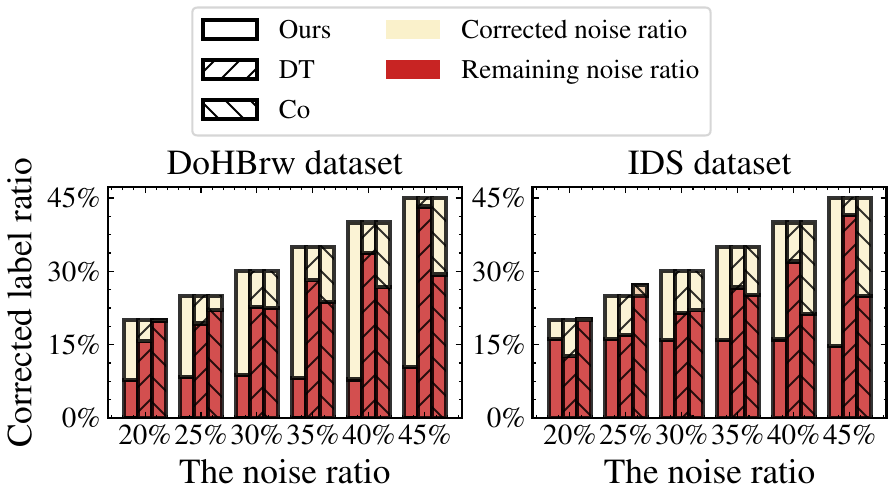}
  \caption{The corrected and remaining noise ratios under different original noise ratios.} 
  \label{fig:label_corrlation}
\vspace{-0.15in}
\end{figure}

\noindent \textbf{Label Correction.} 
Now we compare our label correction module with two SOTA label noise correction methods \textsf{DT} and \textsf{Co}. For a noisy training set, we use our module and these two methods to correct the label noises and calculate the remaining noise ratio and the proportion of corrected noises (\ie the number of corrected noises divided by the total number of noises) as the evaluation metrics. The training size is set to 500 and the noise ratios are consistent with the settings in \S~\ref{subsec:detect-performance-results}. Recall that the IDS and IDS/DoHBrw datasets share the same training set, we only perform label correction experiments on the DoHBrw and IDS datasets.

The label noise correction results are shown in Figure~\ref{fig:label_corrlation}. Our module can reduce the original noise ratio to lower levels. It can reduce the noise ratio to 8.54\% and 15.81\% on the DoHBrw and IDS datasets, respectively. Our module is also stable at correcting noise. The standard deviations of the remaining noise ratios corresponding to our module are only 0.92\% and 0.52\% on the DoHBrw and IDS datasets, respectively. In contrast, the noise correction performance of the other two methods degrades as the original noise ratio increases. When the original noise ratio is 45\%, the proportion of noises corrected by our module is 76.8\% on the DoHBrw dataset, which is 19.97 times higher than \textsf{DT} and 2.21 times higher than \textsf{Co}; our noises correction proportion on the IDS dataset is 67.4\%, which is 8.92 times higher than \textsf{DT} and is 1.52 times higher than \textsf{Co}. We notice the performance of all methods on the IDS dataset is inferior to that on the DoHBrw dataset, mainly due to the difference between the two datasets. 
In summary, our label correction module can effectively correct the incorrect labels under different noise rates, and its performance outperforms the SOTA methods.

\begin{table}[ht]
\centering
\small
\renewcommand{\arraystretch}{1} 
\caption{The detection performance after data augmentation. P, R, and F represent precision, recall, and F1 score.}
\label{tab:data_augmentation}
\resizebox{.48\textwidth}{!}{
\begin{tabular}{c|c|ccccc}
\toprule
Dataset & Metrics & \textsf{NONE} & \textsf{ODDS} & \textsf{SMOTE} & \textsf{GAN} & \textsf{Ours} \\ \midrule

\multirow{3}{*}{DoHBrw} 
& P & .88 $\pm$ .11 & .86 $\pm$ .03 & .82 $\pm$ .10 & .85 $\pm$ .09 & \textbf{.88} $\pm$ .03 \\ 
& R & .93 $\pm$ .06 & .93 $\pm$ .02 & .95 $\pm$ .02 & .93 $\pm$ .04 & \textbf{.98} $\pm$ .01 \\ 
& F & .90 $\pm$ .07 & .89 $\pm$ .01 & .87 $\pm$ .06 & .88 $\pm$ .07 & \textbf{.93} $\pm$ .01 \\ \midrule

\multirow{3}{*}{IDS} 
& P & .67 $\pm$ .15 & .44 $\pm$ .04 & .46 $\pm$ .06 & .48 $\pm$ .04 & \textbf{.69} $\pm$ .12 \\ 
& R & .86 $\pm$ .04 & .89 $\pm$ .03 & .89 $\pm$ .03 & .87 $\pm$ .04 & \textbf{.89} $\pm$ .02 \\ 
& F & .77 $\pm$ .06 & .74 $\pm$ .09 & .59 $\pm$ .03 & .60 $\pm$ .05 & \textbf{.77} $\pm$ .06 \\ \midrule

\multirow{3}{*}{IDS/DoHBrw} 
& P & .54 $\pm$ .14 & .34 $\pm$ .10 & .37 $\pm$ .08 & .36 $\pm$ .06 & \textbf{.56} $\pm$ .02 \\ 
& R & .75 $\pm$ .09 & .93 $\pm$ .02 & .92 $\pm$ .00 & .87 $\pm$ .04 & \textbf{.94} $\pm$ .01 \\ 
& F & .55 $\pm$ .09 & .49 $\pm$ .11 & .53 $\pm$ .07 & .51 $\pm$ .05 & \textbf{.70} $\pm$ .02 \\ \bottomrule
\end{tabular}}
\end{table}

\noindent \textbf{Data Augmentation.} 
We compare our data augmentation module with \textsf{NONE} (refers to perform detection without data augmentation), \textsf{ODDS}~\cite{jan2020odds}, \textsf{SMOTE}~\cite{chawla2002smote}, and the vanilla \textsf{GAN}~\cite{goodfellow2014GAN} model. Each method (except for \textsf{NONE}) generates new training data, combines it with the original training data, and trains the final detection model in our system. We obtain the detection performance on the testing set as the metric to measure each data augmentation method.
We exclude the label noises for fair comparisons, so the training set does not contain label noises and the training and testing sets follow the settings in \S~\ref{subsec:detect-performance-results}.

We show the detection performance when the training size is 500 in Table~\ref{tab:data_augmentation}. Our data augmentation module can achieve the highest F1 score on each dataset, namely,  0.93, 0.77, and 0.70 on the DoHBrw, IDS, and IDS/DoHBrw datasets, respectively. It achieves average improvements of 5.1\%, 22.3\%, and 34.9\% over other data augmentation methods. More crucially, compared with the no augmentation strategy, our module improves both the precision and recall scores. This means our module can identify more unseen malicious testing samples without generating more false alarms. On the IDS/DoHBrw dataset where the difference between the training and testing malicious data is most obvious, our module significantly improves the recall from 0.75 to about 0.94 and the precision from 0.54 to about 0.56. In contrast, the recall improvements of other augmentation approaches are all at the expense of serious precision degradation. 
More seriously, in some cases, the detection performance of \textsf{ODDS}, \textsf{GAN}, and \textsf{SMOTE} is worse than that of the no augmentation strategy. This is because the new training data synthesized by these baselines are not effective at improving the precision or recall scores. \qing{We also observe that a small number of normal samples exhibiting infrequent normal behaviors can lead to lower precision in some cases. This is likely because these samples are located outside dense distribution regions of the normal data and are falsely identified as malicious. Besides, the benefits of the augmentation method vary on different datasets due to the different packet types present in each dataset. For example, the IDS dataset generated by hundreds of hosts contains a wider variety of packet types than the DoH dataset which mainly contains DNS requests. The increased diversity of packet types can lead to more false alarms and lower precisions.}

We supplement the experiment results with the training size of 250 and 1000 in Appendix~\ref{app:data_agumentation} and our module also achieves the best F1 score in these settings. In summary, our data augmentation module can effectively improve the detection capability of ML models built upon limited training data against evolving malware samples.

\subsection{Parameter Deep Dive}
\label{sec:evaluation:parameter}

\begin{figure}[t]
  \centering
  \includegraphics[width=0.85\linewidth]{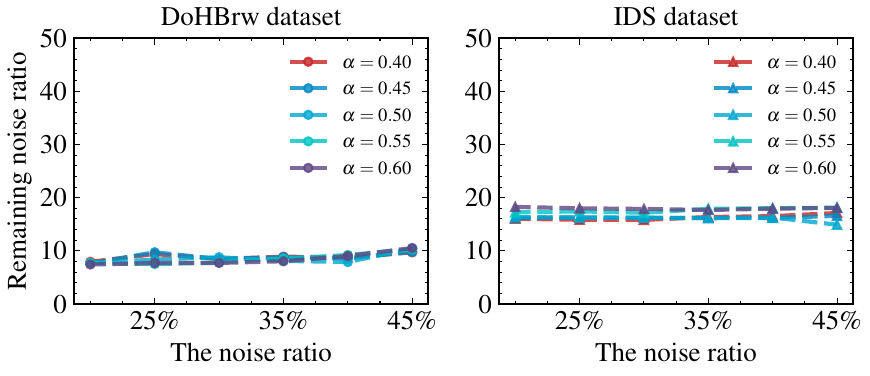}
  \caption{The remaining noise ratios of various filtering ratios $\alpha$.}
  \label{fig:filtering_ratio}
\vspace{-0.15in}
\end{figure}

\noindent \textbf{Filtering Ratio $\bm{\alpha}$.}
We evaluate the impacts of different filtering ratio $\alpha$ values. Recall that $\alpha$ controls the proportion of high-confidence normal samples filtered at the first step of label correction (see \S~\ref{subsec:method-label_correction}). We set five different $\alpha$ values  (0.40, 0.45, 0.50, 0.55, and 0.60). Then, we perform label correction experiments with different noise ratios. Specifically, we set the training size to 500 and make the noise ratio in the range of 20\% to 45\%. Figure~\ref{fig:filtering_ratio} shows the remaining noise ratio results. Our label correction module can always achieve low remaining noise ratios (\ie good label correction results) under different $\alpha$ settings. In particular, the average remaining noise ratios on the DoHBrw and IDS datasets are 8.63\% and 16.84\%, respectively. Moreover, the variance of the results regarding different $\alpha$ settings is relatively low (0.90\% and 0.78\% on the DoHBrw and the IDS dataset, respectively), demonstrating the stability of our label correction module.

\begin{figure}[t]
  \centering
  \includegraphics[width=0.85\linewidth]{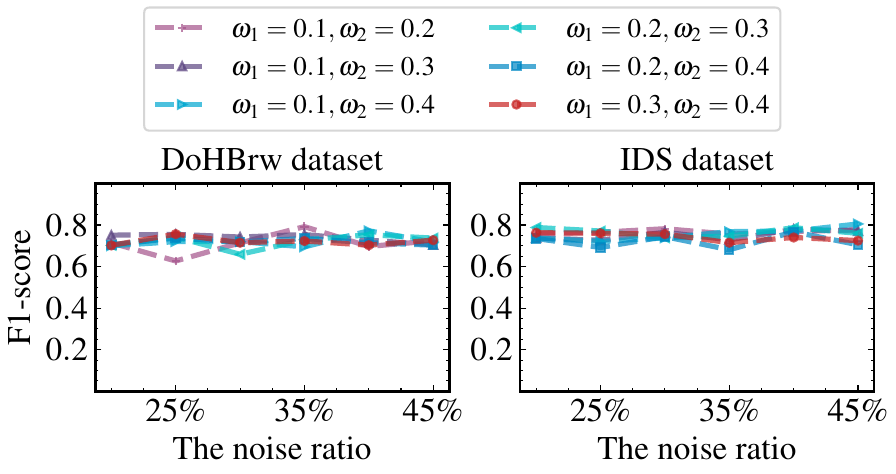}
  \caption{The F1 score of different thresholds $\omega_1, \omega_2$.}
  \label{fig:boundary}
\end{figure}

\begin{figure}[t]
  \centering
  \includegraphics[width=0.85\linewidth]{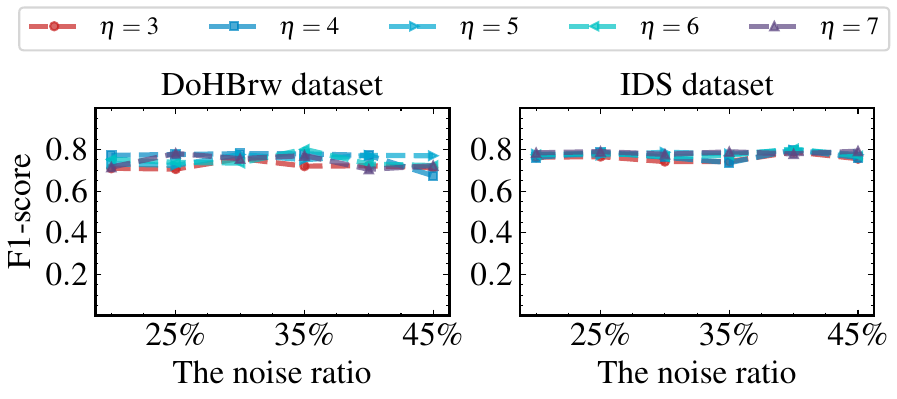}
  \caption{The F1 score of different number of GAN models $\eta$.}
  \label{fig:multiplier}
\vspace{-0.15in}
\end{figure}

\noindent \textbf{Thresholds $\bm{\omega_1}$, $\bm{\omega_2}$ and $\bm{\omega_3}$.} We also evaluate the impacts of different thresholds $\omega_1$, $\omega_2$, and $\omega_3$ values. These thresholds control the size and location of target regions for generated samples (see \S~\ref{subsec:method-data_augmentation}). We use $\omega_1=x$ to denote the value of $\omega_1$ as the $(100 \cdot x)$th percentile of the generated samples, where $x \in [0,1]$, and we use similar notations to present the values of $\omega_2$ and $\omega_3$. 
Besides, we make $\omega_3=\omega_2 + 0.1$ to ensure the new normal samples are generated near the boundary controlled by $\omega_2$. Thus, we only set different values for $\omega_1$ and $\omega_2$. Also, we set $\omega_1$ and $\omega_2$ to less than 0.5 to avoid samples generated inside the region of normal samples, which violates our augmentation strategy. In the experiments, the training size is also set to 500 and the noise ratio is in the range of 20\% to 45\%. We show the F1 scores of our system in Figure~\ref{fig:boundary}. It can be seen that our system achieves relatively stable detection performance when these thresholds vary. For instance, when the noise ratio is 45\%, the standard deviations of F1 scores corresponding to different $\omega_1$ and $\omega_2$ combinations are only 0.025 and 0.016 on the DoHBrw and IDS datasets, respectively.

\noindent\textbf{Number of GAN models $\bm{\eta}$.}
We evaluate the impacts of different $\eta$ values, which represent the number of GANs trained independently for data augmentation. We set the training size to 500 and the noise ratio in the range of 20\% to 45\%. Figure~\ref{fig:multiplier} shows the detection performance of our system under different $\eta$ values. The changing $\eta$ values slightly impact the detection performance. When the noise ratio is 30\%, the standard deviations of F1 scores corresponding to different $\eta$ values on the two datasets are only 0.016 and 0.015. Also, the detection performance improves slightly as $\eta$ increases. In particular, when $\eta$ increases from 3 to 7, the average F1 scores on the DoHBrw and IDS datasets improve by 2.8\% and 3.4\%, respectively. It validates that more independent GAN models can enhance the diversity of the generated training samples, leading to better detection performance. Thus, we can choose either a large $\eta$ value for better detection performance or a moderate $\eta$ value to balance the detection performance and training overhead. 
The overall performance of \ours is not sensitive to parameter choices. Its good detection capability is attributed to our design, rather than specific parameters.

\begin{figure}[t]
  \centering
  \includegraphics[width=0.7\linewidth]{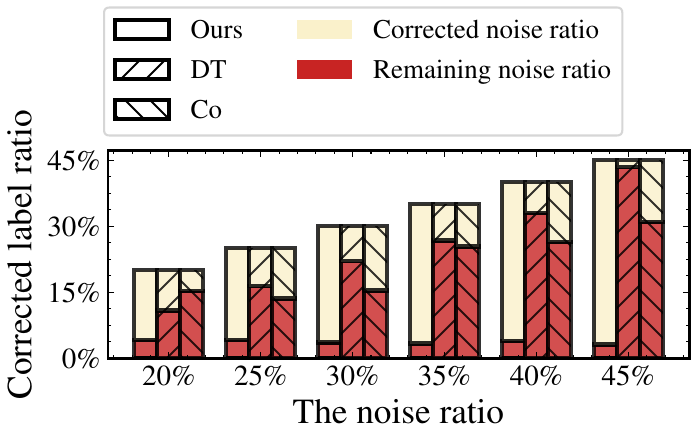}
  \caption{The corrected and remaining noise ratios under different original noise ratios on the real-world dataset.}
  \label{fig:real_label_corrlation}
  \vspace{-0.1in}
\end{figure}

\begin{table}[ht]
\centering
\small
\renewcommand{\arraystretch}{1} 
\caption{The detection performance after data augmentation. P, R, and F represent precision, recall, and F1 score.}
\label{tab:real_data_augmentation}
\resizebox{.48\textwidth}{!}{
\begin{tabular}{c|c|ccccc}
\toprule
Training size & Metrics & \textsf{NONE} & \textsf{ODDS} & \textsf{SMOTE} & \textsf{GAN} & \textsf{Ours} \\ \midrule

\multirow{3}{*}{250} 
& P & .44 $\pm$ .22 & .38 $\pm$ .04 & .45 $\pm$ .06 & .50 $\pm$ .03 & \textbf{.64} $\pm$ .12 \\ 
& R & .14 $\pm$ .02 & .24 $\pm$ .01 & .24 $\pm$ .02 & .14 $\pm$ .01 & \textbf{.90} $\pm$ .11 \\ 
& F & .16 $\pm$ .03 & .22 $\pm$ .02 & .34 $\pm$ .02 & .17 $\pm$ .02 & \textbf{.74} $\pm$ .09 \\ \midrule

\multirow{3}{*}{500} 
& P & .43 $\pm$ .16 & .45 $\pm$ .22 & .27 $\pm$ .06 & .58 $\pm$ .01 & \textbf{.64} $\pm$ .06 \\ 
& R & .21 $\pm$ .02 & .35 $\pm$ .09 & .71 $\pm$ .02 & .34 $\pm$ .02 & \textbf{.96} $\pm$ .05 \\ 
& F & .26 $\pm$ .02 & .32 $\pm$ .13 & .34 $\pm$ .01 & .37 $\pm$ .02 & \textbf{.77} $\pm$ .05 \\ \midrule

\multirow{3}{*}{1000} 
& P & .79 $\pm$ .10 & .69 $\pm$ .14 & .52 $\pm$ .12 & .61 $\pm$ .24 & \textbf{.80} $\pm$ .08 \\ 
& R & .38 $\pm$ .03 & .49 $\pm$ .03 & .64 $\pm$ .22 & .34 $\pm$ .08 & \textbf{.96} $\pm$ .05 \\ 
& F & .47 $\pm$ .04 & .54 $\pm$ .06 & .53 $\pm$ .05 & .38 $\pm$ .01 & \textbf{.87} $\pm$ .04 \\ \bottomrule
\end{tabular}}
\end{table}

\subsection{Real-World Experiments}
\label{subsec:real-world_performance}

To evaluate \ours in real-world cyberspace, we collaborate with a network security enterprise to obtain a large amount of real network data, including both benign and malicious samples. In particular, the dataset is collected on the Internet by the enterprise within its service area. All the traffic samples are labeled with high confidence based on a considerable quantity of threat intelligence collected by the enterprise and further calibrated by experts. In total, we obtain over 2.9M benign and 790K malicious encrypted traffic flows with timestamps ranging from Nov. 2017 to Feb. 2021.

We use the traffic data collected in 2017 as the training set and the rest as the testing set. Then we follow the settings in \S~\ref{subsec:detect-performance-results} to conduct experiments. We supplement the detection performance on the real-world dataset under all settings in Appendix~\ref{app:real_world}.
Overall, our system outperforms all baselines with non-trivial margins in nearly all cases. When the training size is 500, our system achieves the best F1 score of about 0.773, improving existing methods from 89.2\% to 445.5\%, at an average of 272.5\%. 
We also evaluate our label noise correction module on this real-world dataset and show the results in Figure~\ref{fig:real_label_corrlation}. Specifically, our label noise correction module achieves a steadily low remaining noise ratio (less than 4.3\%) given different noise ratios in the training data and the peak ratio for noise removal is 93\%. Besides, we further evaluate our data augmentation module on this real-world dataset, and the results are shown in Table~\ref{tab:real_data_augmentation}. Compared with other augmentation approaches, the new training data synthesized by our module brings in the largest detection performance improvement, especially in the recall. Specifically, the recall score achieved by our data augmentation module outperforms that of other augmentation approaches by around 238.8\% (on average). Meanwhile, the precision score of our module is also higher than all baselines. It shows our data augmentation module enables ML-based detectors to detect more realistic malicious samples unseen in the training data.

\section{Discussion}
\label{sec:discussion}

\noindent\textbf{Extreme Label Noises.} 
The performance of our system will degrade inevitably when more than 50\% of the data is mislabeled (worse than random labeling). However, benefiting from our dedicated design, our system is still more robust than other baselines in such extreme cases. For instance, on the IDS dataset, when the training size is 500 and the label noise ratio is set to 60\%, 75\%, and 90\%, the F1 score of our system still remains at about 0.752, 0.637, and 0.447, respectively. Meanwhile, the performance of \textsf{DT+ODDS+FS} and \textsf{DT+ODDS+ETA} dramatically reduces to about (0.009, 0.010), (0.016, 0.017), and (0.003, 0.000), respectively. Besides, such extreme cases are rare in practice because we can pre-process the collected traffic data to reduce the label noise ratio to a reasonable level, \eg less than 50\%. For instance, we can filter the collected malicious encrypted traffic through the Alexa Top list and only preserve the normal encrypted traffic communicated with well-known benign servers, to reduce the incorrectly labeled traffic mixed in.

\noindent\textbf{Training Overhead.}
Typically, deep learning models involve large training time requirements, yet it is not a critical concern for our system. Since the low-quality training set for our system is usually of limited size, the deep learning models used by our system can be trained fast. For instance, when the number of training samples is 500, the feature extraction module, the label noise correction module, and the data augmentation module consume about 4032, 7.5, and 156 seconds, respectively. Besides, some components in our system, \eg the ensemble classifiers for label noise correction and the multiple GAN models for data augmentation, can be trained in parallel to boost efficiency. Even if the training size becomes larger, we can enhance our system with the recent efficiency improvement techniques, like over-specification~\cite{livni2014computational} and dropout training~\cite{srivastava2014dropout}. 

\noindent\textbf{Performance under Long-term Deployment.}
\qing{We observe that, over the long term, the deployment of our framework will lead to a gradual increase in both the diversity and the volume of normal traffic. This will result in more normal testing traffic exhibiting unseen normal behaviors. These new normal traffic may be detected as malicious.}
A simple yet effective solution is re-collecting the training set to re-train the whole system periodically or when observing obvious performance degradation. As we have evaluated, the training overhead of our system is small (4032, 7.5, and 156 seconds on our three modules, respectively). \qing{We can also cluster these detected malicious samples based on their temporal intervals and filter out the isolated false alarms~\cite{fu2023falsealarm}.} 
Besides, the fresh diverse normal data collected under long-term deployment may affect the effectiveness of our label correction module.
To handle this issue, we can pre-process the normal training data via clustering, \ie aggregating the normal data sharing similar behavior into the same training set, and then correct the label noises in each set individually. We leave detailed exploration in this regard to future work.

\noindent\textbf{Evading Detection.}
Sophisticated attackers may try to bypass our system. One common strategy is generating adversarial examples by adding well-crafted perturbation in the encrypted network traffic of malware. 
Our system has built-in designs to handle adversarial examples. In particular, unlike traditional deep learning models designed for end-to-end tasks, our system uses deep generative models and GAN models to improve the quality of training data \emph{before} training the subsequent ML-based detector. This creates an extra barrier for the adversary to construct adversarial examples for these models. 
In addition, existing art has demonstrated that both label correction and data augmentation can improve the robustness of learning-based models under different attack scenarios~\cite{pang2020tale,rebuffi2021data}. Thus, the ML-based detector in our system is less affected by adversarial examples. 
The second strategy is to mimic the normal network traffic behavior. The pattern/behavior of network flows fundamentally reflects their intent. It is unclear to what extent is this strategy able to retain attack effectiveness while exhibiting identical flow distributions as the benign traffic. We leave exploration for both strategies to future work.
\section{Related Work}
\label{sec:related_work}

\noindent \textbf{Traditional Malicious Traffic Detection.} Identifying the malicious network traffic of malware has been extensively studied. 
\qing{Traditional arts~\cite{bartos2016optimized, jan2020odds, nelms2013execscent, paxson1999bro, roesch1999snort, wang2017detecting} mainly focus on malicious traffic in plain text like URLs and HTTP requests, such that they typically extract attack signatures or features from traffic payloads.} 
For instance, the Execscent method~\cite{nelms2013execscent} extracted five types of features from the URL and HTTP header information and detected malware communication traffic through a template-matching approach. Wang et al.~\cite{wang2017detecting} extracted N-gram semantic features from HTTP headers and selected the most essential features by applying the chi-square test. Besides, to detect the malicious network requests launched by bot malware, Jan et al.~\cite{jan2020odds} convert network traffic into feature vectors based on the frequency of the URL, Referer, and Browser version information. However, these traditional methods are unable to detect encrypted malicious traffic.

\noindent \textbf{Encrypted Malicious Traffic Detection.} 
\qing{Recently a series of malicious encrypted traffic detection methods~\cite{anderson2016cisco, anderson2017cisco, fu2022encrypted, lin2022bert, liu2018mampf, liu2019fsnet, tegeler2012botfinder, wang2016trafficav, xie2023rosetta, zhang2018HoMonit, zhao2022mt, zheng2020learning} have been developed.}
Most of them share a similar design that profiles the behavior of encrypted traffic via manual feature engineering or automatic feature extraction and builds an ML-based detector. For example, the ETA method~\cite{anderson2017cisco} extracted 386 statistical features from the packet length information and TLS handshake metadata and then trained a random forest-based detector. FS-Net~\cite{liu2019fsnet} converted encrypted traffic into sequences of packet lengths and then utilized a novel deep-learning model to classify the traffic into different application categories. However, the effectiveness of these methods relies on a high-quality training set, where the training data is sufficient and correctly labeled. 
\qing{ET-BERT~\cite{lin2022bert} and MT-Flowformer~\cite{zhao2022mt} classify encrypted network traffic under a limited amount of labeled training data by transferring knowledge from large-scale unlabeled training data.}
However, it is time-consuming to collect such training data, which also increases the possibility of privacy leakage from training data. \qing{The detailed comparison of malicious traffic detection methods can be found in Appendix~\ref{app:related_work}.}

\section{Conclusion}

We develop an encrypted malicious traffic system \ours  
that fully utilizes the different distributions of normal and
malicious traffic data in the feature space to augment new data for model training. To the best of our knowledge, \ours is the first malware traffic detection system that simultaneously overcomes the challenges of training data insufficiency and label noises. We implement \ours and perform extensive evaluations based on two public datasets. The experimental results show average improvements of 352.6\%, 284.3\%, and 214.9\% over the SOTA methods on three datasets. We experiment \ours with the dataset from a security enterprise and it effectively achieves malicious traffic detection with the best F1 score of 0.773 and on average 272.5\% improvements of the F1 score over the existing methods.

\section*{Acknowledgements}
\qing{The research is supported in part by the National Key R\&D Project of China under Grant 2021ZD0110502, NSFC under Grant 62132011 and 62221003, China National Funds for Distinguished Young Scientists under Grant 61825204, and the Beijing Outstanding Young Scientist Program under Grant BJJWZYJH01201910003011. Kun Sun's work is supported in part by ONR Grant N00014-23-1-2122. Qi Li is the corresponding author of this paper.} 

\bibliographystyle{IEEEtranS}
\bibliography{references}

\appendix

\subsection{The Detection Performance of All Methods}
\label{app:all_performance}
We evaluate the detection performance of all methods when the training sizes are 250 and 1000 and the label noise changes from 20\% to 45\%. We show the results of all methods on the DoHBrw, IDS, and IDS/DoHBrw dataset in Table~\ref{tab:detection_f1_DoHBrw}, Table~\ref{tab:detection_f1_IM} and Table~\ref{tab:detection_f1_IDS_DoHBrw}, respectively. In general, our system achieves the best F1 score in almost all cases. In particular, when the training size is 250 and the noise ratio is 45\%, the F1 score of our system is 0.59, 0.80, and 0.82 on the DoHBrw, IDS, and IDS/DoHBrw datasets, respectively. This shows an average improvement of 222.5\%, 1330.0\%, and 1523.6\% over all baselines. The reason why our system outperforms other methods is that our system implements label noise correction and three different strategies of data augmentation for more accurate and robust malicious traffic detection.

\begin{table*}[]
    \centering
    \caption{The F1 score of each method on the DoHBrw dataset.}
    \label{tab:detection_f1_DoHBrw}
    \resizebox{\textwidth}{!}{
        \begin{tabular}{c|c|cccccc|cc|c}
            \toprule
            Training size & Noise Ratio &  \textsf{FS} & \textsf{Co+FS} & \textsf{DT+FS} & \textsf{ODDS+FS} & \textsf{SMOTE+FS} & \textsf{DT+ODDS+FS} & \textsf{ETA} & \textsf{DT+ODDS+ETA} & \textsf{Ours} \\
            \midrule 
            \multirow{6}{*}{250} 
 & 20\% & .24 $\pm$ .07 & .24 $\pm$ .04 & .53 $\pm$ .02 & .52 $\pm$ .03 & .16 $\pm$ .00 & .30 $\pm$ .17 & .62 $\pm$ .26 & .37 $\pm$ .07 & \textbf{.74} $\pm$ .03 \\
 & 25\% & .20 $\pm$ .03 & .20 $\pm$ .05 & .48 $\pm$ .02 & .47 $\pm$ .02 & .16 $\pm$ .00 & .52 $\pm$ .07 & .53 $\pm$ .28 & .37 $\pm$ .09 & \textbf{.83} $\pm$ .04 \\
 & 30\% & .18 $\pm$ .01 & .20 $\pm$ .05 & .41 $\pm$ .01 & .41 $\pm$ .04 & .16 $\pm$ .00 & .33 $\pm$ .10 & .56 $\pm$ .27 & .44 $\pm$ .21 & \textbf{.69} $\pm$ .01 \\
 & 35\% & .16 $\pm$ .01 & .20 $\pm$ .04 & .35 $\pm$ .02 & .32 $\pm$ .02 & .16 $\pm$ .00 & .25 $\pm$ .08 & .41 $\pm$ .23 & .29 $\pm$ .08 & \textbf{.68} $\pm$ .01 \\
 & 40\% & .15 $\pm$ .00 & .18 $\pm$ .04 & .27 $\pm$ .02 & .27 $\pm$ .02 & .16 $\pm$ .00 & .30 $\pm$ .02 & .47 $\pm$ .21 & .26 $\pm$ .06 & \textbf{.54} $\pm$ .06 \\
 & 45\% & .14 $\pm$ .00 & .15 $\pm$ .07 & .19 $\pm$ .00 & .18 $\pm$ .02 & .16 $\pm$ .00 & .17 $\pm$ .00 & .32 $\pm$ .25 & .26 $\pm$ .22 & \textbf{.59} $\pm$ .03 \\
            \midrule
            \multirow{6}{*}{1000} 
 & 20\% & .44 $\pm$ .01 & .20 $\pm$ .09 & .66 $\pm$ .01 & .63 $\pm$ .06 & .17 $\pm$ .00 & .57 $\pm$ .15 & .44 $\pm$ .20 & .36 $\pm$ .17 & .58 $\pm$ .03 \\
 & 25\% & .35 $\pm$ .01 & .28 $\pm$ .06 & .62 $\pm$ .02 & .64 $\pm$ .02 & .17 $\pm$ .00 & .62 $\pm$ .04 & .55 $\pm$ .24 & .34 $\pm$ .03 & .58 $\pm$ .01 \\
 & 30\% & .29 $\pm$ .02 & .27 $\pm$ .05 & .54 $\pm$ .02 & .56 $\pm$ .05 & .16 $\pm$ .00 & .57 $\pm$ .05 & .39 $\pm$ .22 & .29 $\pm$ .06 & \textbf{.64} $\pm$ .01 \\
 & 35\% & .25 $\pm$ .03 & .26 $\pm$ .04 & .38 $\pm$ .05 & .46 $\pm$ .02 & .16 $\pm$ .00 & .49 $\pm$ .04 & .57 $\pm$ .32 & .28 $\pm$ .11 & \textbf{.62} $\pm$ .02 \\
 & 40\% & .20 $\pm$ .01 & .22 $\pm$ .06 & .30 $\pm$ .01 & .32 $\pm$ .02 & .16 $\pm$ .00 & .28 $\pm$ .09 & .41 $\pm$ .30 & .28 $\pm$ .14 & \textbf{.62} $\pm$ .02 \\
 & 45\% & .15 $\pm$ .00 & .24 $\pm$ .05 & .20 $\pm$ .02 & .18 $\pm$ .05 & .16 $\pm$ .00 & .20 $\pm$ .02 & .46 $\pm$ .31 & .30 $\pm$ .13 & \textbf{.65} $\pm$ .02 \\
            \bottomrule
        \end{tabular}
    }
\end{table*}
\begin{table*}[]
    \centering
    \caption{The F1 score of each method on the IDS dataset.}
    \label{tab:detection_f1_IM}
    \resizebox{\textwidth}{!}{
    \begin{tabular}{c|c|cccccc|cc|c}
    \toprule
    Training size & Noise Ratio &  \textsf{FS} & \textsf{Co+FS} & \textsf{DT+FS} & \textsf{ODDS+FS} & \textsf{SMOTE+FS} & \textsf{DT+ODDS+FS} & \textsf{ETA} & \textsf{DT+ODDS+ETA} & \textsf{Ours} \\
    \midrule 
    \multirow{6}{*}{250} 
     & 20\% & .35 $\pm$ .02 & .73 $\pm$ .10 & .44 $\pm$ .02 & .45 $\pm$ .05 & .16 $\pm$ .00 & .48 $\pm$ .05 & .70 $\pm$ .16 & .67 $\pm$ .13 & \textbf{.79} $\pm$ .03 \\
 & 25\% & .28 $\pm$ .03 & .70 $\pm$ .10 & .44 $\pm$ .01 & .42 $\pm$ .05 & .16 $\pm$ .00 & .46 $\pm$ .06 & .60 $\pm$ .24 & .61 $\pm$ .20 & \textbf{.81} $\pm$ .01 \\
 & 30\% & .26 $\pm$ .04 & .65 $\pm$ .18 & .06 $\pm$ .00 & .45 $\pm$ .04 & .16 $\pm$ .00 & .15 $\pm$ .00 & .57 $\pm$ .10 & .52 $\pm$ .14 & \textbf{.75} $\pm$ .03 \\
 & 35\% & .19 $\pm$ .02 & .61 $\pm$ .17 & .34 $\pm$ .01 & .34 $\pm$ .05 & .16 $\pm$ .00 & .38 $\pm$ .02 & .39 $\pm$ .22 & .41 $\pm$ .23 & \textbf{.80} $\pm$ .01 \\
 & 40\% & .15 $\pm$ .01 & .34 $\pm$ .15 & .25 $\pm$ .02 & .24 $\pm$ .02 & .16 $\pm$ .00 & .21 $\pm$ .04 & .38 $\pm$ .12 & .40 $\pm$ .15 & \textbf{.79} $\pm$ .02 \\
 & 45\% & .13 $\pm$ .01 & .25 $\pm$ .19 & .00 $\pm$ .00 & .19 $\pm$ .02 & .16 $\pm$ .00 & .08 $\pm$ .07 & .26 $\pm$ .15 & .29 $\pm$ .13 & \textbf{.80} $\pm$ .01 \\
    \midrule
    \multirow{6}{*}{1000}
 & 20\% & .49 $\pm$ .04 & .77 $\pm$ .04 & .59 $\pm$ .04 & .60 $\pm$ .04 & .17 $\pm$ .00 & .62 $\pm$ .02 & .83 $\pm$ .14 & .83 $\pm$ .13 & .77 $\pm$ .00 \\
 & 25\% & .39 $\pm$ .02 & .73 $\pm$ .05 & .56 $\pm$ .06 & .61 $\pm$ .05 & .16 $\pm$ .00 & .53 $\pm$ .07 & .78 $\pm$ .23 & .75 $\pm$ .27 & \textbf{.78} $\pm$ .01 \\
 & 30\% & .35 $\pm$ .00 & .72 $\pm$ .07 & .50 $\pm$ .03 & .50 $\pm$ .08 & .16 $\pm$ .00 & .54 $\pm$ .04 & .65 $\pm$ .27 & .68 $\pm$ .23 & \textbf{.77} $\pm$ .01 \\
 & 35\% & .29 $\pm$ .00 & .67 $\pm$ .09 & .45 $\pm$ .02 & .43 $\pm$ .07 & .16 $\pm$ .00 & .51 $\pm$ .12 & .61 $\pm$ .31 & .63 $\pm$ .31 & \textbf{.75} $\pm$ .03 \\
 & 40\% & .23 $\pm$ .00 & .63 $\pm$ .12 & .37 $\pm$ .01 & .34 $\pm$ .06 & .16 $\pm$ .00 & .39 $\pm$ .01 & .45 $\pm$ .18 & .42 $\pm$ .20 & \textbf{.77} $\pm$ .01 \\
 & 45\% & .16 $\pm$ .01 & .39 $\pm$ .13 & .22 $\pm$ .01 & .25 $\pm$ .04 & .16 $\pm$ .00 & .26 $\pm$ .02 & .40 $\pm$ .22 & .38 $\pm$ .21 & \textbf{.78} $\pm$ .00 \\
    \bottomrule
\end{tabular}
}
\end{table*}

\begin{table*}[]
    \centering
    \caption{The F1 score of each method on the IDS/DoHBrw dataset.}
    \label{tab:detection_f1_IDS_DoHBrw}
    \resizebox{\textwidth}{!}{
        \begin{tabular}{c|c|cccccc|cc|c}
            \toprule
            Training size & Noise Ratio &  \textsf{FS} & \textsf{Co+FS} & \textsf{DT+FS} & \textsf{ODDS+FS} & \textsf{SMOTE+FS} & \textsf{DT+ODDS+FS} & \textsf{ETA} & \textsf{DT+ODDS+ETA} & \textsf{Ours} \\
            \midrule 
            \multirow{6}{*}{250} 
 & 20\% & .33 $\pm$ .03 & .73 $\pm$ .10 & .43 $\pm$ .01 & .53 $\pm$ .03 & .16 $\pm$ .00 & .47 $\pm$ .02 & .70 $\pm$ .23 & .62 $\pm$ .15 & \textbf{.81} $\pm$ .07 \\
 & 25\% & .19 $\pm$ .01 & .70 $\pm$ .10 & .42 $\pm$ .03 & .40 $\pm$ .04 & .16 $\pm$ .00 & .48 $\pm$ .04 & .62 $\pm$ .19 & .57 $\pm$ .20 & \textbf{.81} $\pm$ .09 \\
 & 30\% & .16 $\pm$ .00 & .65 $\pm$ .18 & .08 $\pm$ .04 & .42 $\pm$ .03 & .16 $\pm$ .00 & .15 $\pm$ .00 & .58 $\pm$ .11 & .57 $\pm$ .15 & \textbf{.81} $\pm$ .08 \\
 & 35\% & .16 $\pm$ .01 & .61 $\pm$ .17 & .34 $\pm$ .02 & .36 $\pm$ .01 & .16 $\pm$ .00 & .39 $\pm$ .02 & .43 $\pm$ .22 & .34 $\pm$ .14 & \textbf{.81} $\pm$ .10 \\
 & 40\% & .13 $\pm$ .01 & .34 $\pm$ .15 & .16 $\pm$ .02 & .24 $\pm$ .02 & .16 $\pm$ .00 & .17 $\pm$ .03 & .37 $\pm$ .16 & .38 $\pm$ .12 & \textbf{.81} $\pm$ .08 \\
 & 45\% & .08 $\pm$ .00 & .25 $\pm$ .19 & .00 $\pm$ .00 & .11 $\pm$ .03 & .16 $\pm$ .00 & .05 $\pm$ .05 & .30 $\pm$ .21 & .31 $\pm$ .17 & \textbf{.82} $\pm$ .07 \\
    \midrule
            \multirow{6}{*}{1000} 
  & 20\% & .43 $\pm$ .03 & .77 $\pm$ .04 & .61 $\pm$ .03 & .59 $\pm$ .04 & .17 $\pm$ .00 & .61 $\pm$ .06 & .87 $\pm$ .14 & .79 $\pm$ .14 & .77 $\pm$ .07 \\
 & 25\% & .28 $\pm$ .03 & .73 $\pm$ .05 & .59 $\pm$ .02 & .56 $\pm$ .09 & .16 $\pm$ .00 & .57 $\pm$ .09 & .80 $\pm$ .24 & .79 $\pm$ .17 & \textbf{.83} $\pm$ .04 \\
 & 30\% & .23 $\pm$ .00 & .72 $\pm$ .07 & .51 $\pm$ .03 & .49 $\pm$ .04 & .16 $\pm$ .00 & .52 $\pm$ .08 & .68 $\pm$ .24 & .72 $\pm$ .19 & \textbf{.75} $\pm$ .09 \\
 & 35\% & .19 $\pm$ .00 & .67 $\pm$ .09 & .45 $\pm$ .04 & .44 $\pm$ .07 & .16 $\pm$ .00 & .39 $\pm$ .14 & .62 $\pm$ .27 & .57 $\pm$ .28 & \textbf{.81} $\pm$ .04 \\
 & 40\% & .13 $\pm$ .00 & .63 $\pm$ .12 & .33 $\pm$ .03 & .36 $\pm$ .06 & .16 $\pm$ .00 & .34 $\pm$ .02 & .48 $\pm$ .20 & .43 $\pm$ .18 & \textbf{.79} $\pm$ .08 \\
 & 45\% & .11 $\pm$ .00 & .39 $\pm$ .13 & .22 $\pm$ .03 & .19 $\pm$ .03 & .16 $\pm$ .00 & .27 $\pm$ .02 & .30 $\pm$ .16 & .37 $\pm$ .21 & \textbf{.80} $\pm$ .06 \\
           \bottomrule
        \end{tabular}
    }
\end{table*}
\begin{table*}[]
    \centering
    \caption{The F1 score of each method on the real-world dataset.}
    \label{tab:detection_f1_real}
    \resizebox{\textwidth}{!}{
        \begin{tabular}{c|c|cccccc|cc|c}
            \toprule
            Training size & Noise Ratio &  \textsf{FS} & \textsf{Co+FS} & \textsf{DT+FS} & \textsf{ODDS+FS} & \textsf{SMOTE+FS} & \textsf{DT+ODDS+FS} & \textsf{ETA} & \textsf{DT+ODDS+ETA} & \textsf{Ours} \\
            \midrule 
            \multirow{6}{*}{250} 
 & 20\% & .14 $\pm$ .20 & .05 $\pm$ .16 & .32 $\pm$ .28 & .38 $\pm$ .19 & .17 $\pm$ .06 & .31 $\pm$ .25 & .73 $\pm$ .28 & .78 $\pm$ .30 & .72 $\pm$ .04 \\
 & 25\% & .30 $\pm$ .17 & .08 $\pm$ .25 & .26 $\pm$ .22 & .23 $\pm$ .22 & .16 $\pm$ .04 & .22 $\pm$ .18 & .49 $\pm$ .26 & .65 $\pm$ .27 & \textbf{.70} $\pm$ .05 \\
 & 30\% & .24 $\pm$ .13 & .07 $\pm$ .20 & .23 $\pm$ .15 & .29 $\pm$ .15 & .18 $\pm$ .04 & .23 $\pm$ .13 & .64 $\pm$ .34 & .67 $\pm$ .30 & \textbf{.71} $\pm$ .06 \\
 & 35\% & .23 $\pm$ .13 & .25 $\pm$ .31 & .24 $\pm$ .15 & .25 $\pm$ .15 & .18 $\pm$ .02 & .21 $\pm$ .16 & .47 $\pm$ .39 & .49 $\pm$ .38 & \textbf{.67} $\pm$ .04 \\
 & 40\% & .23 $\pm$ .12 & .16 $\pm$ .23 & .18 $\pm$ .13 & .20 $\pm$ .11 & .16 $\pm$ .03 & .22 $\pm$ .13 & .54 $\pm$ .33 & .60 $\pm$ .35 & \textbf{.86} $\pm$ .04 \\
 & 45\% & .20 $\pm$ .09 & .15 $\pm$ .11 & .22 $\pm$ .09 & .18 $\pm$ .09 & .16 $\pm$ .02 & .21 $\pm$ .09 & .38 $\pm$ .32 & .45 $\pm$ .35 & \textbf{.70} $\pm$ .08 \\
            \midrule
            \multirow{6}{*}{500}
 & 20\% & .15 $\pm$ .18 & .15 $\pm$ .15 & .22 $\pm$ .14 & .15 $\pm$ .16 & .15 $\pm$ .05 & .37 $\pm$ .25 & .68 $\pm$ .39 & .68 $\pm$ .39 & \textbf{.72} $\pm$ .05 \\
 & 25\% & .22 $\pm$ .19 & .20 $\pm$ .36 & .15 $\pm$ .18 & .11 $\pm$ .10 & .13 $\pm$ .06 & .16 $\pm$ .20 & .51 $\pm$ .30 & .58 $\pm$ .31 & \textbf{.76} $\pm$ .06 \\
 & 30\% & .24 $\pm$ .17 & .11 $\pm$ .28 & .24 $\pm$ .19 & .23 $\pm$ .16 & .13 $\pm$ .05 & .25 $\pm$ .18 & .48 $\pm$ .30 & .58 $\pm$ .28 & \textbf{.75} $\pm$ .04 \\
 & 35\% & .20 $\pm$ .14 & .19 $\pm$ .38 & .25 $\pm$ .18 & .19 $\pm$ .20 & .15 $\pm$ .03 & .19 $\pm$ .16 & .31 $\pm$ .34 & .47 $\pm$ .32 & \textbf{.71} $\pm$ .02 \\
 & 40\% & .18 $\pm$ .09 & .17 $\pm$ .29 & .19 $\pm$ .14 & .28 $\pm$ .19 & .11 $\pm$ .06 & .18 $\pm$ .16 & .36 $\pm$ .33 & .45 $\pm$ .35 & \textbf{.76} $\pm$ .04 \\
 & 45\% & .20 $\pm$ .09 & .12 $\pm$ .15 & .17 $\pm$ .10 & .17 $\pm$ .09 & .13 $\pm$ .05 & .17 $\pm$ .11 & .16 $\pm$ .25 & .17 $\pm$ .27 & \textbf{.77} $\pm$ .06 \\
            \midrule
            \multirow{6}{*}{1000} 
 & 20\% & .30 $\pm$ .27 & .00 $\pm$ .01 & .29 $\pm$ .34 & .27 $\pm$ .23 & .13 $\pm$ .06 & .24 $\pm$ .21 & .73 $\pm$ .31 & .66 $\pm$ .28 & \textbf{.81} $\pm$ .02 \\
 & 25\% & .36 $\pm$ .21 & .00 $\pm$ .00 & .29 $\pm$ .25 & .23 $\pm$ .20 & .15 $\pm$ .04 & .21 $\pm$ .21 & .55 $\pm$ .27 & .44 $\pm$ .26 & \textbf{.64} $\pm$ .05 \\
 & 30\% & .13 $\pm$ .15 & .32 $\pm$ .46 & .18 $\pm$ .15 & .12 $\pm$ .16 & .12 $\pm$ .05 & .16 $\pm$ .14 & .68 $\pm$ .34 & .43 $\pm$ .33 & .61 $\pm$ .01 \\
 & 35\% & .20 $\pm$ .15 & .41 $\pm$ .42 & .28 $\pm$ .21 & .25 $\pm$ .16 & .12 $\pm$ .05 & .30 $\pm$ .19 & .50 $\pm$ .35 & .45 $\pm$ .34 & \textbf{.69} $\pm$ .04 \\
 & 40\% & .29 $\pm$ .14 & .19 $\pm$ .30 & .32 $\pm$ .21 & .24 $\pm$ .18 & .15 $\pm$ .01 & .22 $\pm$ .18 & .35 $\pm$ .33 & .42 $\pm$ .37 & \textbf{.63} $\pm$ .02 \\
 & 45\% & .14 $\pm$ .09 & .04 $\pm$ .07 & .22 $\pm$ .16 & .17 $\pm$ .10 & .14 $\pm$ .02 & .17 $\pm$ .09 & .25 $\pm$ .26 & .35 $\pm$ .30 & \textbf{.68} $\pm$ .02 \\
            \bottomrule
        \end{tabular}
    }
\end{table*}

\subsection{The Comparison of Different Data Augmentation Methods}
\label{app:data_agumentation}
We compare the performance of different data augmentation methods when the training size is set to 250 and 1000. We utilize each method (except for \textsf{NONE}) to generate new training data, combine it with the original training data, and train the final detection model in our system to perform detection. The results on the three datasets are shown in Table~\ref{tab:data_augmentation_DoHBrw}, Table~\ref{tab:data_augmentation_IM} and Table~\ref{tab:data_augmentation_IDS_DoHBrw}. It can be seen that our data augmentation module significantly outperforms other methods. The best F1 score of our module is about 0.95, 0.73, and 0.72 on the three datasets. 

\begin{table*}[ht]
    \centering
    \small
    \renewcommand{\arraystretch}{1.1}
    \caption{The detection performance when using different data augmentation methods on the DoHBrw dataset.}
    \label{tab:data_augmentation_DoHBrw}
    \resizebox{0.4\textwidth}{!}{
        \begin{tabular}{c|c|ccc}
            \toprule
            Training size & Method & Precision & Recall & F1 score \\
            \midrule 
            \multirow{5}{*}{250} 
& \textsf{NONE} & .66 $\pm$ .20 & .91 $\pm$ .04 & .74 $\pm$ .12 \\
& \textsf{ODDS} & .54 $\pm$ .08 & .96 $\pm$ .02 & .69 $\pm$ .06 \\
&\textsf{SMOTE} & .54 $\pm$ .15 & .95 $\pm$ .02 & .68 $\pm$ .12 \\
&  \textsf{GAN} & .58 $\pm$ .20 & .93 $\pm$ .04 & .69 $\pm$ .14 \\
& \textsf{Ours} & \textbf{.77} $\pm$ .04 & \textbf{.97} $\pm$ .01 & \textbf{.86} $\pm$ .02 \\
             \midrule 
            \multirow{5}{*}{1000}
& \textsf{NONE} & .86 $\pm$ .06 & .92 $\pm$ .05 & .89 $\pm$ .01 \\
& \textsf{ODDS} & .69 $\pm$ .16 & .93 $\pm$ .03 & .78 $\pm$ .10 \\
&\textsf{SMOTE} & .61 $\pm$ .11 & .95 $\pm$ .02 & .74 $\pm$ .08 \\
&  \textsf{GAN} & .64 $\pm$ .14 & .95 $\pm$ .02 & .75 $\pm$ .10 \\
& \textsf{Ours} & \textbf{.94} $\pm$ .03 & \textbf{.96} $\pm$ .02 & \textbf{.95} $\pm$ .01 \\
            \bottomrule 
        \end{tabular}
    }
\end{table*}

\begin{table*}[ht]
    \centering
    \small
    \renewcommand{\arraystretch}{1.1}
    \caption{The detection performance when using different data augmentation methods on the IDS dataset.}
    \label{tab:data_augmentation_IM}
    \resizebox{0.4\textwidth}{!}{
        \begin{tabular}{c|c|ccc}
            \toprule
            Training size & Method & Precision & Recall & F1 score\\
            \midrule 
            \multirow{5}{*}{250}
& \textsf{NONE} & .42 $\pm$ .03 & .82 $\pm$ .05 & .55 $\pm$ .01 \\
& \textsf{ODDS} & .37 $\pm$ .04 & .83 $\pm$ .02 & .52 $\pm$ .03 \\
&\textsf{SMOTE} & .21 $\pm$ .01 & .93 $\pm$ .02 & .35 $\pm$ .01 \\
&  \textsf{GAN} & .35 $\pm$ .02 & .86 $\pm$ .03 & .49 $\pm$ .02 \\
& \textsf{Ours} & \textbf{.44} $\pm$ .03 & .92 $\pm$ .01 & \textbf{.60} $\pm$ .03 \\
             \midrule 
            \multirow{5}{*}{1000}
&  \textsf{NONE} & .55 $\pm$ .05 & .89 $\pm$ .01 & .67 $\pm$ .04 \\
&  \textsf{ODDS} & .47 $\pm$ .06 & .89 $\pm$ .02 & .62 $\pm$ .05 \\
& \textsf{SMOTE} & .30 $\pm$ .02 & .95 $\pm$ .00 & .45 $\pm$ .02 \\
&   \textsf{GAN} & .48 $\pm$ .07 & .88 $\pm$ .02 & .62 $\pm$ .06 \\
&  \textsf{Ours} & \textbf{.63} $\pm$ .05 & \textbf{.87} $\pm$ .02 & \textbf{.73} $\pm$ .03 \\
            \bottomrule 
        \end{tabular}
    }
\end{table*}

\begin{table*}[ht]
    \centering
    \small
    \renewcommand{\arraystretch}{1.1}
    \caption{The detection performance when using different data augmentation methods on the IDS/DoHBrw dataset.}
    \label{tab:data_augmentation_IDS_DoHBrw}
    \resizebox{0.4\textwidth}{!}{
        \begin{tabular}{c|c|ccc}
            \toprule
            Training size & Method & Precision & Recall & F1 score\\
            \midrule 
            \multirow{5}{*}{250}
&  \textsf{NONE} & .34 $\pm$ .01 & .83 $\pm$ .04 & .48 $\pm$ .02 \\
&  \textsf{ODDS} & .37 $\pm$ .00 & .90 $\pm$ .01 & .53 $\pm$ .00 \\
& \textsf{SMOTE} & .19 $\pm$ .00 & .93 $\pm$ .01 & .31 $\pm$ .00 \\
&   \textsf{GAN} & .35 $\pm$ .02 & .85 $\pm$ .06 & .50 $\pm$ .02 \\
&  \textsf{Ours} & \textbf{.59} $\pm$ .08 & .92 $\pm$ .00 & \textbf{.72} $\pm$ .05 \\
             \midrule 
            \multirow{5}{*}{1000}
&  \textsf{NONE} & .55 $\pm$ .10 & .79 $\pm$ .13 & .63 $\pm$ .03 \\
&  \textsf{ODDS} & .46 $\pm$ .08 & .83 $\pm$ .08 & .58 $\pm$ .04 \\
& \textsf{SMOTE} & .26 $\pm$ .02 & .93 $\pm$ .02 & .41 $\pm$ .03 \\
&   \textsf{GAN} & .48 $\pm$ .15 & .76 $\pm$ .17 & .55 $\pm$ .02 \\
&  \textsf{Ours} & \textbf{.55} $\pm$ .03 & .87 $\pm$ .02 & \textbf{.67} $\pm$ .01 \\
            \bottomrule 
        \end{tabular}
    }
\end{table*}
\begin{table*}[ht]
    \centering
    \renewcommand{\arraystretch}{1.1} 
    \caption{\qing{The comparison of malicious traffic detection methods. Stats and Metadata mean statistics features and TLS handshake metadata, respectively.}}
    \label{tab:related_work}
    \resizebox{.9\textwidth}{!}{
    \begin{tabular}{c|c|cccc}
         \toprule
         \multicolumn{2}{c|}{Malicious traffic detection methods} & Granularity & Input & Robustness under label noises & Limited training data\\
         \midrule
         \multirow{4}{*}{\makecell{Traditional \\ malicious \\ traffic detection}} &
         \cite{bartos2016optimized, nelms2013execscent, wang2017detecting} & HTTP Request & Payload & \XSolidBrush & \XSolidBrush \\&
         \cite{jan2020odds} & HTTP Request & Payload & \XSolidBrush & \CheckmarkBold \\&
         \cite{paxson1999bro} & Flow & Payload & \XSolidBrush & \XSolidBrush \\ &
         \cite{roesch1999snort} & Packet & Payload & \XSolidBrush & \XSolidBrush
         \\ 
         \midrule
         \multirow{10}{*}{\makecell{Encrypted \\ malicious \\ traffic detection}} &
         \cite{anderson2016cisco} & Flow & Stats \& Metadata & \XSolidBrush & \XSolidBrush \\ &
         \cite{anderson2017cisco} & Flow & Stats \& Packet Sequence \& Metadata & \XSolidBrush & \XSolidBrush \\ &
         \cite{fu2022encrypted} & Flow Set & Metadata & \XSolidBrush & \XSolidBrush \\ &
         \cite{lin2022bert} & Flow & Traffic Raw Byte & \XSolidBrush & \CheckmarkBold \\ &
         \cite{liu2018mampf} & Flow & Packet Sequence \& Metadata & \XSolidBrush & \XSolidBrush \\ &
         \cite{liu2019fsnet, xie2023rosetta} & Flow & Packet Sequence & \XSolidBrush & \XSolidBrush \\ &
         \cite{tegeler2012botfinder, wang2016trafficav} & Flow & Stats & \XSolidBrush & \XSolidBrush \\ &
         \cite{zhang2018HoMonit} & Flow Set & Packet Sequence & \XSolidBrush & \XSolidBrush \\ &
          \cite{zhao2022mt,zheng2020learning} & Flow & Packet Sequence & \XSolidBrush & \CheckmarkBold \\ &
         \ours & Flow & Packet Sequence & \CheckmarkBold & \CheckmarkBold \\
         \bottomrule
    \end{tabular}
    }
\end{table*}

\subsection{The Detection Performance on the Real-World Dataset}
\label{app:real_world}

We conduct experiments on the real-world dataset with various training sizes (250, 500, and 1000) and noise ratio (20\%, 25\%, 30\%, 35\%, 40\%, and 45\%) settings, and the results are shown in Table~\ref{tab:detection_f1_real}. Overall, our system achieves the highest F1 score in almost all cases.

\subsection{More Comparisons with Existing Methods}
\label{app:related_work}
\qing{
Malicious traffic detection has been extensively studied~\cite{anderson2016cisco, anderson2017cisco, bartos2016optimized, fu2022encrypted, jan2020odds, lin2022bert, liu2018mampf, liu2019fsnet, nelms2013execscent, paxson1999bro, roesch1999snort, tegeler2012botfinder, wang2016trafficav, wang2017detecting, xie2023rosetta,  zhang2018HoMonit, zhao2022mt, zheng2020learning}. 
A number of detection methods have been developed. They detect malicious traffic by using different information. Unfortunately, they are unable to detect malicious traffic under limited training data with label noises. The detailed comparisons can be found in Table~\ref{tab:related_work}. 
}

\end{document}